\documentclass[final,3p,times]{elsarticle}

\usepackage{amssymb}

\usepackage{amssymb}
\usepackage[]{amsmath}
\usepackage{graphics}
\usepackage{amssymb}
\usepackage[]{amsmath}
\usepackage{amsthm}
\usepackage{setspace}
\usepackage{epsfig}
\usepackage{subfigure}
\usepackage{cases}
\usepackage{graphicx}

\biboptions{numbers,sort&compress}

\usepackage{amsmath}
\usepackage{times}
\usepackage{anysize}
\marginsize{2.5cm}{2.5cm}{1cm}{2cm}

\selectfont

\begin{document}

\begin{frontmatter}


\title{Breather to the Yajima-Oikawa system}

\author{Junchao Chen \fnref{label11,label1}}
\ead{junchaochen@aliyun.com}
\author{Yong Chen \fnref{label1} }
\ead{ychen@sei.ecnu.edu.cn}
\author{Bao-Feng Feng \fnref{label2} }
{\ead{baofeng.feng@utrgv.edu}}
\author{Ken-ichi Maruno \fnref{label3}}
\ead{kmaruno@waseda.jp}


\cortext[cor1]{Corresponding author.}
\address[label11]{Department of Mathematics, Lishui University, Lishui, 323000, People¡¯s Republic of China}
\address[label1]{ Shanghai Key Laboratory of Trustworthy Computing, East China Normal University, Shanghai, 200062, People's Republic of China}
\address[label2]{Department of Mathematics, The University of Texas Rio Grande Valley, Edinburg,
TX 78539, USA}
\address[label3]{Department of Applied Mathematics, School of Fundamental Science and Engineering,
Waseda University, 3-4-1 Okubo, Shinjuku-ku, Tokyo 169-8555, Japan}

\begin{abstract}
The Yajima-Oikawa (YO) system describes the resonant interaction between long and short waves under certain condition.
In this paper, through the KP hierarchy reduction, we construct the breather solutions for the YO system in one- and two-dimensional cases.
Similar to Akhmediev and Kuznetsov-Ma breather solutions (the wavenumber $k_i\rightarrow {\rm i}k_i$ ) for the nonlinear Schr\"{o}dinger equation,
is shown that the YO system have two kinds of breather solutions with the relations
$p_{2k-1}\rightarrow{\rm i}p_{2k-1}$, $p_{2k}\rightarrow-{\rm i}p_{2k}$, $q_{2k-1}\rightarrow{\rm i}q_{2k-1}$ and $q_{2k}\rightarrow-{\rm i}q_{2k}$, in which the homoclinic orbit and dark soliton solutions are two special cases respectively.
Furthermore, taking the long wave limit, we derive the rational and rational and rational-exp solutions
which contain lump, line rogue wave, soliton and their mixed cases.
By considering the further reduction, such solutions can be reduced to one-dimensional YO system.

\end{abstract}

\end{frontmatter}


\section{Introduction}

Since the soliton concept was first introduced by Zabusky and Kruskal \cite{zabusky1965interaction} in 1965,
the investigation of integrable equations has become an important subject in nonlinear science (see, e.g.
refs. \cite{zakharov1991integrability} and \cite{ablowitzsoliton} and reference therein).
As a result of balance between nonlinearity and dispersion, soltions, which are stable localized
waves, describe a class of undistorted nonlinear wave propagation over large distance.
Apart from solitons, the breather solutions of integrable equations have also attracted more and
more attention because of their generic properties in a wide range of nonlinear systems.

The so-called breather reflects the behavior of the localized
waves which is periodic in time or space and localized in space or time.
The Akhmediev breather solution \cite{akhmediev1986modulation} for the nonlinear Schr\"{o}dinger (NLS) equation is periodic in space and localized in time,
whereas the Kuznetsov-Ma \cite{kuznetsov1977solitons,ma1979perturbed} breather solution is periodic in time and decrease exponentially in space.
The Peregrine breather solution being not periodic in both space and time, also known as rational solution, has been proposed in \cite{peregrine1983water} and it has been considered as special prototype of rogue waves.
As is reported \cite{dysthe1999note,kedziora2012second}, three kinds of breather solutions are related, especially, the Peregrine breathers are some special cases of the Akhmediev and the Kuznetsov-Ma solution.
The homoclinic solutions of NLS equation, with spatially
periodic boundary conditions, are the most common unstable wave packets closely associated
chaos \cite{ablowitz1990homoclinic}. Osborne \cite{osborne2014classification} has investigated an infinite number of other homoclinic
solutions of NLS and showed that they reduce to the above three classical breather
solutions for particular spectral values in the periodic inverse scattering transform.

The one-dimensional Yajima-Oikawa (YO) equation was originally derived to describe
the interaction of Langmuir and sound waves in plasma \cite{yajima1976formation}.
In the literature \cite{benney1977general,djordjevic1977two,grimshaw1977modulation,funakoshi1983resonant},
it and its variants were usually called the long wave-short wave resonance interaction (LSRI) equation,
which describe a resonant interaction when the group velocity of the short wave matches the phase velocity of the long wave.
The one-dimensional YO system was solved exactly by using the inverse scattering method and shown to have N-bright-soliton solutions \cite{yajima1976formation,ma1978complete}.
Ma and Redekopp \cite{ma1979some} have rewritten YO equation into Hirota's bilinear forms to obtain the bright- and
dark-soltion solutions.
The homoclinic breather solutions of the YO equation were obtained by a B\"{a}cklund transformation and the dressing method \cite{wright2006homoclinic} and also were derived through the bilinear method recently \cite{chow2013rogue}.
Then the rogue wave solutions for one-dimensional YO equation were discussed in detail in Ref. \cite{chow2013rogue,chen2014dark,chen2014darboux}.

The two-dimensional YO system for the resonant interaction between a long surface wave and a short internal wave in a two-layer fluid was presented in \cite{grimshaw1977modulation,oikawa1989two}.
Meanwhile, the bright- and dark- soltion solutions are provided in \cite{oikawa1989two} by using Hirota's direct method. The Painlev\'{e} property for the two-dimensional YO system was investigated in \cite{radha2005periodic} and some special solutions such as positons, dromions, instantons and periodic wave solutions were constructed in \cite{radha2005periodic,lai1999wave}.
In this paper, by using the Hirota's bilinear method and the KP hierarchy reduction method,
we will construct the breather solutions for the YO system in one- and two-dimensional case.
Moreover, by taking the long wave limit, we also derive the rational and rational-exp solutions
which exhibit the abundant structures.

\section{Bilinear form and Gram determinant solution}

\subsection{Bilinear form}
In this paper, we consider the one-dimensional YO system:
\begin{eqnarray}
\label{byo-01} &&\textrm{i} S_t -S_{xx} -L S=0,\\
\label{hyo-02} &&L_t-2a L_x+2\sigma(SS^*)_x=0.
\end{eqnarray}
and two-dimensional YO system:
\begin{eqnarray}
\label{byo-03} &&\textrm{i} S_t + \textrm{i} S_y - S_{xx} - L S=0,\\
\label{byo-04} &&L_t-2a L_x+2\sigma(SS^*)_x=0.
\end{eqnarray}

By using the dependent variables transformations
\begin{eqnarray}
\label{byo-05} S=\rho{\rm e}^{-{\rm i}\alpha t}\frac{g}{f},\ \ S^*=\rho{\rm e}^{{\rm i}\alpha t}\frac{g^*}{f},\ \ L=\alpha+2\left( \ln f \right)_{xx},
\end{eqnarray}
Eqs.(\ref{byo-03})-(\ref{byo-04}) become the following bilinear form
\begin{eqnarray}
\label{byo-06} &&[{\rm i}D_t +{\rm i}D_y- D^2_x]g \cdot f=0,\\
\label{byo-07} &&[{\rm i}D_t +{\rm i}D_y+ D^2_x]g^* \cdot f=0,\\
\label{byo-08} &&[D_{x}D_{t} -2aD^2_x -2C]f \cdot f+2\sigma\rho^2g\hat{g}=0, \ \ (C=\sigma\rho^2).
\end{eqnarray}
where the Hirota's bilinear operators $D_x$,$D_y$ and $D_t$ are defined as
\begin{eqnarray*}
D^n_xD^m_yD^l_t(a\cdot b)=\bigg( \frac{\partial}{\partial x} - \frac{\partial}{\partial x'} \bigg)^n
\bigg( \frac{\partial}{\partial y} - \frac{\partial}{\partial y'} \bigg)^m
\bigg( \frac{\partial}{\partial t} - \frac{\partial}{\partial t'} \bigg)^l
a(x,y,t)b(x',y',t')\bigg|_{x=x',y=y',t=t'}.
\end{eqnarray*}
Then, we introduce another group of (2+1)-dimensional bilinear equations:
\begin{eqnarray}
\label{byo-09} &&[{\rm i}D_t +{\rm i}D_y- D^2_x]g \cdot f=0,\\
\label{byo-10} &&[{\rm i}D_t +{\rm i}D_y+ D^2_x]\hat{g} \cdot f=0,\\
\label{byo-11} &&[D_{x}D_{t} -2aD^2_x -2C]f \cdot f+2\sigma\rho^2g\hat{g}=0, \ \ (C=\sigma\rho^2).
\end{eqnarray}
If the solutions $f,g$ and $\hat{g}$ of bilinear equations (\ref{byo-09})-(\ref{byo-10}) further satisfy the conditions,
\begin{eqnarray}
\label{byo-12} f^*=f \ \ \mbox{and}\ \  g^*=\hat{g},
\end{eqnarray}
then these solutions also satisfy the bilinear YO equations (\ref{byo-06})-(\ref{byo-08}).

\subsection{Gram determinant solution for the (2+1)-dimensional system (\ref{byo-09})-(\ref{byo-10})}

\textbf{Lemma 2.1} The following bilinear equations in the KP hierarchy
\begin{eqnarray}
\label{byo-13} && (D_{x_1}D_{x_{-1}}-2)\tau(k) \cdot \tau(k)=-2\tau(k+1)\tau(k-1),\\
\label{byo-14} && (D^2_{x_1}-D_{x_2}+2 \tilde{a} D_{x_1}) \tau(k+1) \cdot \tau(k)=0,
\end{eqnarray}
where $a$ is a complex constant and $k$ is an integer,
have the Gram determinant solution
\begin{eqnarray}
\label{byo-15} \tau(k)=\left| m_{ij}(k) \right|_{1\leq i,j \leq N},
\end{eqnarray}
where the matrix element $m_{ij}(k)$ satisfies
\begin{eqnarray}\label{byo-16}
\nonumber && \partial_{x_1} m_{ij}(k)=\phi_i(k)\psi_j(k),\\
\nonumber && \partial_{x_2} m_{ij}(k,l)=[\partial_{x_1}\phi_i(k)]\psi_j(k)-\phi_i(k)[\partial_{x_1}\psi_j(k)],\\
&& \partial_{x_{-1}} m_{ij}(k)=-\phi_i(k-1)\psi_j(k+1),\\
\nonumber &&  m_{ij}(k+1)=m_{ij}(k)+\phi_i(k,l)\psi_j(k+1),
\end{eqnarray}
and
\begin{eqnarray}\label{byo-17}
\nonumber && \partial_{x_2}\phi_i(k)=\partial^2_{x_1}\phi_i(k),\\
&& \phi_i(k+1)=(\partial_{x_1}-\tilde{a})\phi_i(k),\\
\nonumber && \partial_{x_2}\psi_j(k)=-\partial^2_{x_1}\psi_j(k),\\
\nonumber && \psi_j(k-1)=-(\partial_{x_1}+\tilde{a})\psi_j(k).
\end{eqnarray}

\textbf{Lemma 2.2} The bilinear equations (\ref{byo-13})-(\ref{byo-14}) in the KP hierarchy have the Gram determinant solution as follows
\begin{eqnarray}\label{byo-18}
\tau(k)=\left| m_{ij}(k) \right|_{1\leq i,j \leq N},
\end{eqnarray}
where the matrix element $m_{ij}(k)$ is defined by
\begin{eqnarray*}
&& m_{ij}(k)=c_{ij}+ \frac{1}{p_i+q_j} \phi_i(k)\psi_j(k) ,\\
&& \phi_i(k)=(p_i-\tilde{a})^k (p_i+q_i) \exp(r_i),\\
&& \psi_j(k)=\left(-\frac{1}{q_i+\tilde{a}}\right)^k \exp(s_j),
\end{eqnarray*}
with
\begin{eqnarray*}
&& r_i=\frac{1}{p_i-\tilde{a}}x_{-1}  + p_ix_1 + p^2_i x_2 + r_{i0},\\
&& s_j=\frac{1}{q_j+\tilde{a}}x_{-1}  + q_jx_1 - q^2_j x_2 + s_{j0},
\end{eqnarray*}
where $c_{ij},p_i,q_j,r_{i0}$ and $s_{j0}$ are complex constants.

\emph{Proof}.   It is easy to verify that functions $m_{ij}(k)$, $\phi_i(k)$ and $\psi_j(k)$ satisfy the differential and difference rules (\ref{byo-16})--(\ref{byo-17}).

By introducing independent variables transformations
\begin{eqnarray}\label{byo-19}
x_1=-{\rm i}(x+2at), \ \ x_2= {\rm i}y,\ \ x_{-1}={\rm i} \sigma\rho^2(t-y),
\end{eqnarray}
ie.,
\begin{eqnarray}\label{byo-20}
\partial_{x_1}={\rm i} \partial_x,\ \ \partial_{x_2}= 2{\rm i}a\partial_x -{\rm i}\partial_y -{\rm i} \partial_t, \ \
\partial_{x_{-1}}= \frac{2{\rm i}a}{\sigma\rho^2}\partial_{x}  - \frac{{\rm i}}{\sigma \rho^2} \partial_{t},
\end{eqnarray}
and defining
\begin{eqnarray}\label{byo-21}
f=\tau(0),\ \ g=\tau(1),\ \ \hat{g}=\tau(-1),
\end{eqnarray}
and $\tilde{a}=a$ , the bilinear equations (\ref{byo-13})-(\ref{byo-14}) become (\ref{byo-09})-(\ref{byo-11}).

Thus, Eqs.(\ref{byo-09})-(\ref{byo-11}) have soliton solution given by Gram determinants
\begin{eqnarray}
\label{byo-22}&& f=\Bigg|  \delta_{ij} + \frac{p_i+q_i}{p_i+q_j} \textmd{e}^{\xi_i+\eta_j} \Bigg|_{N\times N}, \\
\label{byo-23}&& g=
\Bigg| \delta_{ij} + \left(-\frac{p_i-a}{q_j+a}\right)  \frac{p_i+q_i}{p_i+q_j} \textmd{e}^{\xi_i+\eta_j} \Bigg|_{N\times N}, \\
\label{byo-24}&& \hat{g}=
\Bigg| \delta_{ij} + \left(-\frac{q_j+a}{p_i-a} \right) \frac{p_i+q_i}{p_i+q_j}  \textmd{e}^{\xi_i+\eta_j}  \Bigg|_{N\times N},
\end{eqnarray}
with
\begin{eqnarray*}
\xi_i=-{\rm i}p_i(x+2at) + {\rm i}p^2_iy +{\rm i} \frac{\sigma\rho^2(t-y)}{p_i-a}+\xi_{i,0},\ \
\eta_i=-{\rm i}q_i(x+2at) - {\rm i}q^2_iy + {\rm i}\frac{\sigma\rho^2(t-y)}{q_i+a}+\eta_{i,0}.
\end{eqnarray*}


In order to satisfy the complex conjugate condition (\ref{byo-12}), we present two different case in the following paper.

\section{Breather-I }
In this section, we give two kind of breather solutions to the original YO system with the completely different complex conjugate form, although the final parameterization process show that one is a special case of another result.
By taking the long wave limit, we also derive the rational and mixed rational-exponential solutions directly.

\subsection{Complex conjugacy}

\subsubsection{Complex conjugation A}

With the purpose of deriving the breather solution to the YO system, we assume that the integer $N$ in (\ref{byo-22})-(\ref{byo-24}) is even,  ie., $N=2M$,
then the solutions $f,g$ and $\hat{g}$ can be rewritten as
\begin{eqnarray}
&& f
=\Delta_0
\Bigg|  \frac{\delta_{ij}(-1)^{i+1}}{(p_i+q_i)\textmd{e}^{\xi_i+\eta_j}} + \frac{(-1)^{i+1} }{p_i+q_j}  \Bigg|_{N\times N}
\equiv \Delta_0\left|  F  \right|= \Delta_0\left|  (F_{ij})_{1 \leq i,j\leq N}  \right|, \\
&& g=
\Delta_0
\Bigg| \frac{\delta_{ij}(-1)^{i+1}}{(p_i+q_i)\textmd{e}^{\xi_i+\eta_j}} + \left(-\frac{p_i-a}{q_j+a}\right)  \frac{(-1)^{i+1}}{p_i+q_j} \Bigg|_{N\times N}
\equiv \Delta_0\left|  G  \right|=\Delta_0\left|  (G_{ij})_{1 \leq i,j\leq N}  \right|, \\
&& \hat{g}=
\Delta_0
\Bigg| \frac{\delta_{ij}(-1)^{i+1}}{(p_i+q_i)\textmd{e}^{\xi_i+\eta_j}} + \left(-\frac{q_i+a}{p_j-a} \right) \frac{(-1)^{i+1}}{p_i+q_j}    \Bigg|_{N\times N}
 \equiv \Delta_0 |  \hat{G}  |=\Delta_0 |  (\hat{G}_{ij})_{1 \leq i,j\leq N}  |,
\end{eqnarray}
with
\begin{eqnarray*}
\Delta_0=\textmd{e}^{\sum^{2M}_{i=1}\xi_i+\eta_i} \prod^{M}_{k=1}[-(p_{2k-1}+q_{2k-1})(p_{2k}+q_{2k})].
\end{eqnarray*}

By taking
\begin{eqnarray}\label{byo-28}
&& p_{2k-1}=-\Omega_k+\frac{\omega_k}{2},\ \ p_{2k}=-\Omega^*_k-\frac{\omega_k}{2},\ \ q_{2k-1}=\Omega_k+\frac{\omega_k}{2},\ \ q_{2k}=\Omega^*_k-\frac{\omega_k}{2},\ \
\end{eqnarray}
and $\xi_{2k-1,0}=\xi_{2k,0} \equiv \xi_{k,0},\ \ \eta_{2k-1,0}=\eta_{2k,0} \equiv \eta_{k,0}$, where $\Omega_k$ are complex parameters and $\omega_k,\xi_{k,0},\eta_{k,0}$ are real parameters for $k=1,2,\cdots,M$,
we have
\begin{eqnarray}
\nonumber && \xi_{2k-1}+\eta_{2k-1}=\xi^*_{2k}+\eta^*_{2k} \\
&&\hspace{2.2cm} = -\textrm{i}w_k x
+ \left[ \frac{4{\rm i}\sigma \rho^2\omega_k}{(2\Omega_k+2a)^2-\omega^2_k} -2{\rm i}\omega_k\Omega_k \right]y
+ [-\frac{4{\rm i}\sigma \rho^2\omega_k}{(2\Omega_k+2a)^2-\omega^2_k}-2{\rm i}a\omega_k] t
+\xi_{k,0}+\eta_{k,0},\ \
\end{eqnarray}
and $\Delta_0$ is real.

Let
\begin{eqnarray}
\tilde{F}=E_0F^TE^T_0,\ \ \tilde{G}=E_0\hat{G}^TE^T_0,
\end{eqnarray}
where $^T$ denotes the transposition, and $E_0$ is an antisymmetric  $2M\times 2M$ matrix:
\begin{eqnarray*}
E_0=\left( \begin {array}{ccccccc}   0 & 1 & 0&0 & \cdots & 0&0\\
-1 & 0 & 0&0 & \cdots & 0&0\\
0&0&0 & 1 &\cdots & 0&0\\
0&0&-1 & 0 &\cdots & 0&0\\
\vdots &\vdots & \vdots&\vdots & \ddots  &  \vdots & \vdots \\
 0&0&0&0&0 &  0 & 1 \\
 0&0&0&0&0 & -1 & 0 \\
 \end {array} \right),
\end{eqnarray*}
one can find
\begin{eqnarray}
\tilde{F}_{ij}=F^*_{ij}, \ \ \tilde{G}_{ij}=G^*_{ij},
\end{eqnarray}
and then
\begin{eqnarray}
|F^*|=|\tilde{F}|=|E_0F^TE^T_0|=|F|,\ \ |G^*|=|\tilde{G}|=|E_0\hat{G}^TE^T_0|=|\hat{G}|.
\end{eqnarray}
Finally, we can get $f^*=f$ and $g^*=\hat{g}$.
That is to say, the solutions $f,g$ and $\hat{g}$ of bilinear equations (\ref{byo-09})-(\ref{byo-10}) satisfy the conditions (\ref{byo-12}).
The above result is summarized as follows:

\textbf{Theorem 3.1} The breather solutions for the two-dimensional YO system are
\begin{eqnarray}\label{byo-33}
 S=\rho{\rm e}^{-{\rm i}\alpha t}\frac{g}{f},\ \ L=\alpha+2\left( \ln f \right)_{xx},
\end{eqnarray}
where $f= \Delta_0|F_{k,l}|$, $g= \Delta_0|G_{k,l}|$ and
$\Delta_0=\textmd{e}^{\sum^{M}_{i=1}\zeta_i+\zeta^*_i} \prod^{M}_{k=1}\omega^2_k$,
and the matrix elements are defined by
\begin{eqnarray*}
&& F_{k,k}= \left( \begin {array}{cc}
\frac{1}{\omega_k {\rm e}^{\zeta_k}} + \frac{1}{\omega_k} &  - \frac{1}{\Omega_k-\Omega^*_k} \\
- \frac{1}{\Omega_k-\Omega^*_k}  &  \frac{1}{\omega_k {\rm e}^{\zeta^*_k}} + \frac{1}{\omega_k}
 \end {array} \right),\\
&& G_{k,k}= \left( \begin {array}{cc}
\frac{1}{\omega_k {\rm e}^{\zeta_k}} + \frac{1}{\omega_k}\frac{\Omega_k+a+ \frac{\omega_k}{2}}{\Omega_k+a- \frac{\omega_k}{2}} &
  - \frac{1}{\Omega_k-\Omega^*_k}\frac{\Omega_k+a- \frac{\omega_k}{2}}{\Omega^*_k+a- \frac{\omega_k}{2}} \\
- \frac{1}{\Omega_k-\Omega^*_k} \frac{\Omega^*_k+a+ \frac{\omega_k}{2}}{\Omega_k+a+ \frac{\omega_k}{2}}  &
 \frac{1}{\omega_k {\rm e}^{\zeta^*_k}} + \frac{1}{\omega_k}
 \frac{\Omega^*_k+a+ \frac{\omega_k}{2}}{\Omega^*_k+a- \frac{\omega_k}{2}}
 \end {array} \right),\\
&& F_{k,l}=\left( \begin {array}{cc}
\frac{1}{-(\Omega_k-\Omega_l)+\frac{\omega_k+\omega_l}{2}} &
\frac{1}{-(\Omega_k-\Omega^*_l)+\frac{\omega_k-\omega_l}{2}}\\
\frac{1}{(\Omega^*_k-\Omega_l)+\frac{\omega_k-\omega_l}{2}} &
\frac{1}{(\Omega^*_k-\Omega_l)+\frac{\omega_k+\omega_l}{2}}
 \end {array} \right),\\
&& G_{k,l}=\left( \begin {array}{cc}
\frac{1}{-(\Omega_k-\Omega_l)+\frac{\omega_k+\omega_l}{2}} \frac{\Omega_k+a -\frac{\omega_k}{2} }{\Omega_l+a +\frac{\omega_l}{2} }  &
\frac{1}{-(\Omega_k-\Omega^*_l)+\frac{\omega_k-\omega_l}{2}} \frac{\Omega_k+a -\frac{\omega_k}{2} }{\Omega^*_l+a -\frac{\omega_l}{2} } \\
\frac{1}{(\Omega^*_k-\Omega_l)+\frac{\omega_k-\omega_l}{2}} \frac{\Omega^*_k+a +\frac{\omega_k}{2} }{\Omega_l+a +\frac{\omega_l}{2} } &
\frac{1}{(\Omega^*_k-\Omega_l)+\frac{\omega_k+\omega_l}{2}} \frac{\Omega^*_k+a +\frac{\omega_k}{2} }{\Omega^*_l+a -\frac{\omega_l}{2} }
 \end {array} \right),
\end{eqnarray*}
with
\begin{eqnarray*}
&& \zeta_{k} = -\textrm{i}\omega_k x
+ \left[ \frac{4{\rm i}\sigma \rho^2\omega_k}{(2\Omega_k+2a)^2-\omega^2_k} -2{\rm i}\omega_k\Omega_k \right]y
+ [-\frac{4{\rm i}\sigma \rho^2\omega_k}{(2\Omega_k+2a)^2-\omega^2_k}-2{\rm i}a\omega_k] t
+\xi_{k,0}+\eta_{k,0}.
\end{eqnarray*}
where $\Omega_k$ are complex parameters and $\omega_k,\xi_{k,0},\eta_{k,0}$ are real parameters for $k=1,2,\cdots,M$.
When these parameters satisfy the constraint conditions:
\begin{eqnarray}\label{byo-34}
\frac{2\sigma \rho^2}{(2\Omega_k+2a)^2-\omega^2_k} -\Omega_k =0,
\end{eqnarray}
the solutions (\ref{byo-33}) are breather solutions for the one-dimensional YO system.

\subsubsection{Complex conjugation B}

Similarly, considering that the integer $N$ in (\ref{byo-22})-(\ref{byo-24}) is even,  ie., $N=2M$,
then the solutions $f,g$ and $\hat{g}$ can also be rewritten as
\begin{eqnarray}
&& f
=\Delta_0
\Bigg|  \frac{\delta_{ij}}{(p_i+q_i)\textmd{e}^{\xi_i+\eta_j}} + \frac{1 }{p_i+q_j}  \Bigg|_{N\times N}
\equiv \Delta_0\left|  F  \right|= \Delta_0\left|  (F_{ij})_{1 \leq i,j\leq N}  \right|, \\
&& g=
\Delta_0
\Bigg| \frac{\delta_{ij}}{(p_i+q_i)\textmd{e}^{\xi_i+\eta_j}} + \left(-\frac{p_i-a}{q_j+a}\right)  \frac{1}{p_i+q_j} \Bigg|_{N\times N}
\equiv \Delta_0\left|  G  \right|=\Delta_0\left|  (G_{ij})_{1 \leq i,j\leq N}  \right|, \\
&& \hat{g}=
\Delta_0
\Bigg| \frac{\delta_{ij}}{(p_i+q_i)\textmd{e}^{\xi_i+\eta_j}} + \left(-\frac{q_i+a}{p_j-a} \right) \frac{1}{p_i+q_j}    \Bigg|_{N\times N}
 \equiv \Delta_0 |  \hat{G}  |=\Delta_0 |  (\hat{G}_{ij})_{1 \leq i,j\leq N}  |,
\end{eqnarray}
with
\begin{eqnarray*}
&& \Delta_0=\textmd{e}^{\sum^{2M}_{i=1}\xi_i+\eta_i} \prod^{M}_{k=1}[(p_{2k-1}+q_{2k-1})(p_{2k}+q_{2k})],
\end{eqnarray*}

By taking
\begin{eqnarray}\label{byo-38}
&& p_{2k-1}=a+\Omega_k,\ \ p_{2k}=a-\Omega_k,\ \ q_{2k-1}=-a+\Omega^*_k,\ \ q_{2k}=-a-\Omega^*_k,\ \
\end{eqnarray}
and $\xi_{2k-1,0}=\xi_{2k,0} \equiv \xi_{k,0},\ \ \eta_{2k-1,0}=\eta_{2k,0} \equiv \eta_{k,0}$, where $\Omega_k$ are complex parameters and $\omega_k,\xi_{k,0},\eta_{k,0}$ are real parameters for $k=1,2,\cdots,M$,
we have
\begin{eqnarray}
\nonumber && \xi_{2k-1}+\eta_{2k-1}=\xi^*_{2k}+\eta^*_{2k}\\
 &&\hspace{2.2cm} = (\Omega_k+\Omega^*_k)\left[-\textrm{i} x + (-\frac{\textrm{i}\sigma\rho^2}{\Omega_k\Omega^*_k}-\textrm{i}(\Omega^*_k-\Omega_k-2a) )y +(\frac{\textrm{i}\sigma\rho^2}{\Omega_k\Omega^*_k}-2\textrm{i}a  )t    \right]
 +\xi_{k,0}+\eta_{k,0}.\ \ \ \
\end{eqnarray}
Further,
\begin{eqnarray}
F_{ij}=F^*_{ji},\ \ G^*_{ij}=\hat{G}_{ji},
\end{eqnarray}
which leads to $f^*=f$ and $g^*=\hat{g}$. So one can obtain the following theorem :

\textbf{Theorem 3.2} The breather solutions for the two-dimensional YO system are
\begin{eqnarray}\label{byo-41}
 S=\rho{\rm e}^{-{\rm i}\alpha t}\frac{g}{f},\ \ L=\alpha+2\left( \ln f \right)_{xx},
\end{eqnarray}
where $f= \Delta_0|F_{k,l}|$, $g= \Delta_0|G_{k,l}|$ and
$\Delta_0=\textmd{e}^{\sum^{M}_{i=1}\zeta_i+\zeta^*_i} \prod^{M}_{k=1}[-(\Omega_k+\Omega^*_k)^2]$,
and the matrix elements are defined by
\begin{eqnarray*}
&& F_{k,k}= \left( \begin {array}{cc}
\frac{1}{(\Omega_k+\Omega^*_k) {\rm e}^{\zeta_k}} + \frac{1}{\Omega_k+\Omega^*_k} &   \frac{1}{\Omega_k-\Omega^*_k} \\
- \frac{1}{\Omega_k-\Omega^*_k}  &  -\frac{1}{(\Omega_k+\Omega^*_k) {\rm e}^{\zeta^*_k}} - \frac{1}{\Omega_k+\Omega^*_k}
 \end {array} \right),\\
&& G_{k,k}= \left( \begin {array}{cc}
\frac{1}{(\Omega_k+\Omega^*_k) {\rm e}^{\zeta_k}} - \frac{1}{\Omega_k+\Omega^*_k} \frac{\Omega_k}{\Omega^*_k} &   \frac{1}{\Omega_k-\Omega^*_k} \frac{\Omega_k}{\Omega^*_k}  \\
- \frac{1}{\Omega_k-\Omega^*_k}  \frac{\Omega_k}{\Omega^*_k}  &
 -\frac{1}{(\Omega_k+\Omega^*_k) {\rm e}^{\zeta^*_k}} + \frac{1}{\Omega_k+\Omega^*_k} \frac{\Omega_k}{\Omega^*_k}
 \end {array} \right),\\
&& F_{k,l}=\left( \begin {array}{cc}
\frac{1}{\Omega_k+\Omega^*_l} &
\frac{1}{\Omega_k-\Omega^*_l}\\
-\frac{1}{\Omega_k-\Omega^*_l} &
-\frac{1}{\Omega_k+\Omega^*_l}
 \end {array} \right),\\
&& G_{k,l}=\left( \begin {array}{cc}
-\frac{1}{\Omega_k+\Omega^*_l}\frac{\Omega_k}{\Omega^*_l} &
\frac{1}{\Omega_k-\Omega^*_l} \frac{\Omega_k}{\Omega^*_l}\\
-\frac{1}{\Omega_k-\Omega^*_l} \frac{\Omega_k}{\Omega^*_l}&
\frac{1}{\Omega_k+\Omega^*_l} \frac{\Omega_k}{\Omega^*_l}
 \end {array} \right),
\end{eqnarray*}
with
\begin{eqnarray*}
&& \zeta_{k} = (\Omega_k+\Omega^*_k)\left[-\textrm{i} x + [-\frac{\textrm{i}\sigma\rho^2}{\Omega_k\Omega^*_k}-\textrm{i}(\Omega^*_k-\Omega_k-2a) ]y +(\frac{\textrm{i}\sigma\rho^2}{\Omega_k\Omega^*_k}-2\textrm{i}a  )t    \right]
 +\zeta_{k,0}.
\end{eqnarray*}

\subsubsection{Parameterization of A and B}
For the two different complex conjugate way, we show that the breather solutions (\ref{byo-41}) is just a special case of the former ones (\ref{byo-33}) through the parameterization process.

For (\ref{byo-28}),
if we let
\begin{eqnarray}
\Omega_k=-a-\frac{\omega_k}{2}\frac{1}{\tanh(A_k+{\rm i}\phi_k)},
\end{eqnarray}
then
\begin{eqnarray}\label{byo-43}
 && \nonumber  p_{2k-1}= a+ \frac{\omega_k}{1-{\rm e}^{-2A_k-2{\rm i}\phi_k}},\ \
   p_{2k}= a- \frac{\omega_k}{1-{\rm e}^{2A_k-2{\rm i}\phi_k}},\\
 &&  q_{2k-1}= -a+ \frac{\omega_k}{1-{\rm e}^{2A_k+2{\rm i}\phi_k}},\ \
   q_{2k}= -a- \frac{\omega_k}{1-{\rm e}^{-2A_k+2{\rm i}\phi_k}}.\ \
\end{eqnarray}

For (\ref{byo-38}),
if we let
\begin{eqnarray}
&&\Omega_k=\frac{\omega_k}{2} \left(1-\frac{{\rm i}}{\tan\phi_k } \right),
\end{eqnarray}
then
\begin{eqnarray}\label{byo-45}
&&\nonumber p_{2k-1}= a+ \frac{\omega_k}{1-{\rm e}^{-2{\rm i}\phi_k}},\ \
   p_{2k}= a- \frac{\omega_k}{1-{\rm e}^{-2{\rm i}\phi_k}},\\
&& q_{2k-1}= -a+ \frac{\omega_k}{1-{\rm e}^{2{\rm i}\phi_k}},\ \
   q_{2k}= -a- \frac{\omega_k}{1-{\rm e}^{2{\rm i}\phi_k}}.\ \
\end{eqnarray}
Here $A_k$ and $\phi_k$ are real parameters for $k=1,2,\cdots,M$.

By comparing  the new parametric expressions (\ref{byo-43}) and (\ref{byo-45}),
it is easy to find that (\ref{byo-43}) are reduced to (\ref{byo-45}) just by taking $A=0$.
In fact, for the two-dimensional YO system, the breather solutions (\ref{byo-41}), in which the frequency parameters are pure imaginary, are derived by using the same way of deriving the general dark soliton solutions.
The breather solutions (\ref{byo-33}) with the complex frequency parameters are viewed as generalization of (\ref{byo-41}).
The similar extension of the breather solution exists for the focusing NLS equation, which represents the general breather.
This generalization is of importance: when the two-dimensional YO system is reduced to the one-dimensional case,
one can't obtain its breather solutions from the latter solutions.
This fact has been also pointed out by Chow \cite{chow2013rogue} for $M=1$.

\subsubsection{Breather-I solution for $M=1$ and $M=2$ }

When $M=1$, the one-breather solutions for the two-dimensional YO system are written as
\begin{eqnarray}
\label{byo-46} && S=\rho{\rm e}^{-{\rm i}\alpha t}\frac{g}{f},\ \ L=\alpha+2\left( \ln f \right)_{xx},\\
\label{byo-47} &&f= 1+{\rm e}^{\zeta_1}+{\rm e}^{\zeta^*_1}
+A_{12}{\rm e}^{\zeta_1+\zeta^*_1},\\
\label{byo-48} &&g= 1
+\frac{2\Omega_1+2a-\omega_1}{2\Omega_1+2a+\omega_1}{\rm e}^{\zeta_1}
+\frac{2\Omega^*_1+2a+\omega_1}{2\Omega^*_1+2a-\omega_1}{\rm e}^{\zeta^*_1}
+A_{12} \frac{(2\Omega_1+2a-\omega_1)(2\Omega^*_1+2a+\omega_1)}{(2\Omega_1+2a+\omega_1)(2\Omega^*_1+2a-\omega_1)} {\rm e}^{\zeta_1+\zeta^*_1},
\end{eqnarray}
with $A_{12}=1-\frac{\omega^2_1}{(\Omega_1-\Omega^*_1)^2}$ and $\zeta_{1}
= -\textrm{i}w_1 x
+ \left[ \frac{4{\rm i}\sigma \rho^2\omega_1}{(2\Omega_1+2a)^2-\omega^2_1} -2{\rm i}\omega_1\Omega_1 \right]y
+ [-\frac{4{\rm i}\sigma \rho^2\omega_1}{(2\Omega_1+2a)^2-\omega^2_1}-2{\rm i}a\omega_1] t
+\zeta_{1,0}$.

Let $\Omega_1=\Omega_{1R}+{\rm i}\Omega_{1I}$, then $\zeta_1=\zeta_{1R}+{\rm i}\zeta_{1I}$,
\begin{eqnarray}
&& \zeta_{1R}=2\omega_1\left[\frac{16\sigma \rho^2\Omega_{1I} (\Omega_{1R}+a) }{\Theta^2_1+64\Omega^2_{1I} (\Omega_{1R}+a)^2} +\Omega_{1I}\right]y -\frac{32\sigma \rho^2\omega_1\Omega_{1I} (\Omega_{1R}+a) }{\Theta^2_1+64\Omega^2_{1I} (\Omega_{1R}+a)^2}t+\zeta_{1,0},\\
&& \zeta_{1I}=-w_1 x+2\omega_1\left[\frac{ 2\sigma \rho^2\Theta_1 }{\Theta^2_1+64\Omega^2_{1I} (\Omega_{1R}+a)^2} -\Omega_{1R}\right]y-2\omega_1[\frac{ 2\sigma \rho^2\Theta_1 }{\Theta^2_1+64\Omega^2_{1I} (\Omega_{1R}+a)^2} +a ]t,
\end{eqnarray}
where $ \Theta_1=4(\Omega_{1R}+a)^2-4\Omega^2_{1I} -\omega^2_1$.

It is straightforward to show that $S\rightarrow\rho{\rm e}^{-{\rm i}\alpha t}$, $L\rightarrow \alpha$ as $\zeta_{1R}\rightarrow -\infty$ and
$S\rightarrow\rho{\rm e}^{-{\rm i}\alpha t} \frac{4(\Omega_{1R}+a)^2-(\omega_1-{\rm i}\Omega_{1I})^2}{4(\Omega_{1R}+a)^2-(\omega_1+{\rm i}\Omega_{1I})^2} $, $L\rightarrow \alpha$ as $\zeta_{1R}\rightarrow \infty$.
Thus this breather solution is localized along the direction of the line $\zeta_{1R}=0$ and periodic along the direction of the line $\zeta_{1I}=0$. This breather behavior is controlled by the complex parameter $\Omega_1$ and the real spatial frequency $\omega_1$. Fig. 1 illustrates the behavior of this breather solution only at the fixed time $t=0$.

\begin{figure*}[!htbp]
\centering
{\includegraphics[height=1.2in,width=3in]{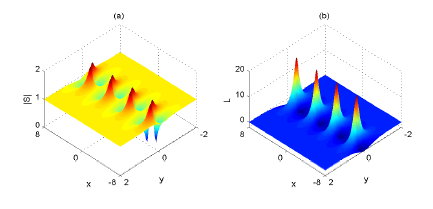}}
\caption{  The one general breather for two-dimensional YO system defined by Eqs. (\ref{byo-46})--(\ref{byo-48}) with the parameters $\sigma=\rho=a=1$, $\zeta_1=0$, $\Omega_1=\frac{1}{20}+{\rm i}$ and $\omega_1=\frac{3}{2}$ at the time $t=0$. }
\end{figure*}

\begin{figure*}[!htbp]
\centering
{\includegraphics[height=2.1in,width=4.3in]{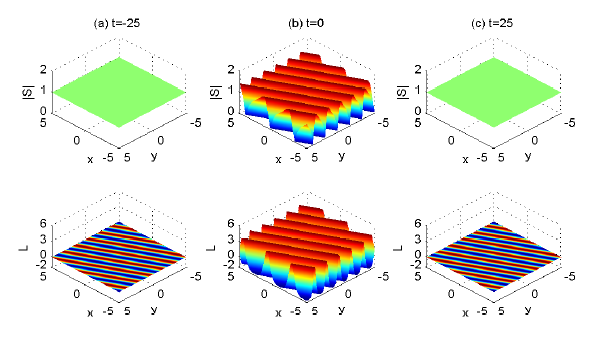}}
\caption{  The first homoclinic orbit for two-dimensional YO system defined by Eqs. (\ref{byo-46})--(\ref{byo-48}) and (\ref{byo-51}) with the parameters $\sigma=\rho=1$, $a=\zeta_1=0$,  $\Omega_1=-1+\frac{1}{2}{\rm i}$ and $\omega_1=\sqrt{3}$. }
\end{figure*}

Note that if $\frac{\partial}{\partial_y}\zeta_{1R}=0$, ie.,
\begin{eqnarray}\label{byo-51}
\frac{16\sigma \rho^2\Omega_{1I} (\Omega_{1R}+a) }{\Theta^2_1+64\Omega^2_{1I} (\Omega_{1R}+a)^2} +\Omega_{1I}=0,
\end{eqnarray}
the localized behavior only occurs in time.
This special case represents the homoclinic orbit for the two-dimensional YO system.
One example of this solution, the first homoclinic orbit is shown in Fig.2.
 If $\Omega_1=\frac{1}{2p}(q+{\rm i}\Omega),\omega_1=p,\sigma=1,\rho=1,a=0,{\rm e}^{\xi_{1,0}}=b_4,\eta_{1,0}=\gamma$ and $S=u,L=-v,\alpha=a$, and conditions $2(\Omega^2_1+\Omega^{*2}_1)-\omega^2_1=0, (\Omega_1+\Omega^*_1)(\Omega_1-\Omega^*_1)^2=2 $  one can get Sheng's result \cite{shen2008shen}.

More generally, defining $\Omega_k=\Omega_{kR}+{\rm i}\Omega_{kI}$ and $\Theta_k=4(\Omega_{kR}+a)^2-4\Omega^2_{kI} -\omega^2_k$ for $k=1,2,\cdots,M$, when the constraint conditions
\begin{eqnarray}\label{byo-511}
\frac{16\sigma \rho^2\Omega_{kI} (\Omega_{kR}+a) }{\Theta^2_k+64\Omega^2_{kI} (\Omega_{kR}+a)^2} +\Omega_{kI}=0,
\end{eqnarray}
are satisfied, the combination homoclinic solutions with $M$ modes are constructed directly.

For the one-dimensional case, the constraint conditions (\ref{byo-34}) become $\frac{2\sigma \rho^2}{(2\Omega_1+2a)^2-\omega^2_1} -\Omega_1 =0$ and the solution corresponds to the first homoclinic orbit. This kind of solution was also reported by Chow \cite{chow2013rogue} ($ a=-\frac{\Delta}{2},\rho=\rho_0,\sigma=\frac{\sigma}{2},\omega_1=p$ and $\Omega_1=-a-\frac{{\rm i}\Omega^*}{2p}$).

\begin{figure*}[!htbp]
\centering
{\includegraphics[height=2.1in,width=4.3in]{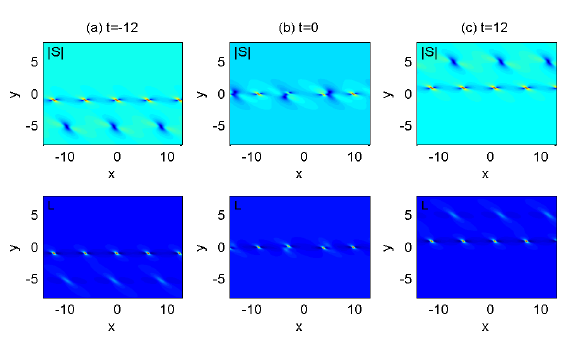}}
\caption{  The two general breather for two-dimensional YO system defined by Eqs. (\ref{byo-52})--(\ref{byo-54}) with the parameters $\sigma=\rho=1$, $a=\zeta_1=\zeta_2=0$,  $\Omega_1=\frac{2}{5}+\frac{4}{3}{\rm i},\Omega_2=1+\frac{1}{2}{\rm i}$, $\omega_1=1$ and $\omega_2=\frac{2}{3}$. \label{two-gen-breather-2d}}
\end{figure*}

\begin{figure*}[!htbp]
\centering
{\includegraphics[height=2.1in,width=4.3in]{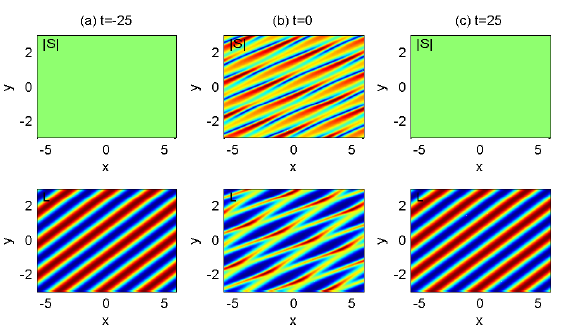}}
\caption{  The second homoclinic orbit for two-dimensional YO system defined by Eqs. (\ref{byo-52})--(\ref{byo-54}) and (\ref{byo-511})$(k=1,2)$ with the parameters $\sigma=\rho=1$, $a=\zeta_1=\zeta_2=0$,  $\Omega_1=-1+\frac{1}{2}{\rm i},\Omega_2=-\frac{25}{16}+\frac{2}{5}{\rm i}$, $\omega_1=\sqrt{3}$ and $\omega_2=\frac{\sqrt{14601}}{40}$. \label{two-homclinic-2d} }
\end{figure*}

\begin{figure*}[!htbp]
\centering
{\includegraphics[height=2.1in,width=4.3in]{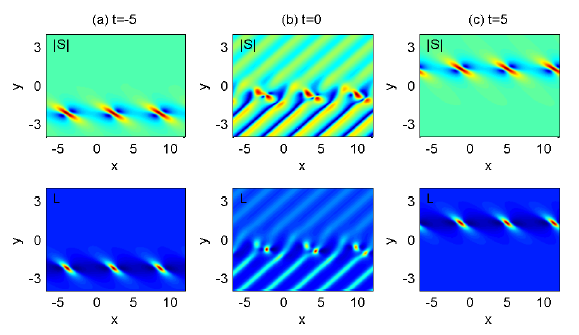}}
\caption{  The mixed solution consisting of one-breather and one homoclinic orbit for two-dimensional YO system defined by Eqs. (\ref{byo-52})--(\ref{byo-54}) and (\ref{byo-511})$(k=1)$ with the parameters $\sigma=\rho=1$, $a=\zeta_1=\zeta_2=0$,  $\Omega_1=-1+\frac{1}{2}{\rm i},\Omega_2=\frac{2}{5}+\frac{2}{3}{\rm i}$, $\omega_1=\sqrt{3}$ and $\omega_2=1$. \label{two-breather-homoclinic-2d} }
\end{figure*}

When $M=2$, the two-breather solutions for the two-dimensional YO system are written as
\begin{eqnarray}
\label{byo-52} && S=\rho{\rm e}^{-{\rm i}\alpha t}\frac{g}{f},\ \ L=\alpha+2\left( \ln f \right)_{xx},\\
\label{byo-53} &&\nonumber  f= 1+{\rm e}^{\zeta_1} +{\rm e}^{\zeta_2} +{\rm e}^{\zeta^*_1} +{\rm e}^{\zeta^*_2}\\
&&\nonumber \ \ \ \ \ \ +K_1{\rm e}^{\zeta_1+\zeta^*_1}+K_2{\rm e}^{\zeta_2+\zeta^*_2}
+ K_{12}{\rm e}^{\zeta_1+\zeta_2}+ K^*_{12}{\rm e}^{\zeta^*_1+\zeta^*_2}
+ \tilde{K}_{12}{\rm e}^{\zeta_1+\zeta^*_2}+ \tilde{K}^*_{12}{\rm e}^{\zeta^*_1+\zeta_2}\\
&&\nonumber \ \ \ \ \ \ +K_1K_{12}\tilde{K}^*_{12} {\rm e}^{\zeta_1+\zeta^*_1+\zeta_2}
+ K_1K^*_{12}\tilde{K}_{12} {\rm e}^{\zeta_1+\zeta^*_1+\zeta^*_2}
+ K_2K_{12}\tilde{K}_{12}{\rm e}^{\zeta_1+\zeta_2+\zeta^*_2}
+ K_2K^*_{12}\tilde{K}^*_{12}{\rm e}^{\zeta^*_1+\zeta_2+\zeta^*_2}\\
&& \ \ \ \ \ \ + K_1K_2 K_{12}K^*_{12}\tilde{K}_{12}\tilde{K}^*_{12} {\rm e}^{\zeta_1+\zeta^*_1+\zeta_2+\zeta^*_2},\\
\label{byo-54} &&\nonumber g= 1+H_1{\rm e}^{\zeta_1} +H_2{\rm e}^{\zeta_2}
+\frac{1}{H^*_1}{\rm e}^{\zeta^*_1} +\frac{1}{H^*_2}{\rm e}^{\zeta^*_2}\\
&&\nonumber \ \ \ \ \ \
+\frac{H_1}{H^*_1}K_1{\rm e}^{\zeta_1+\zeta^*_1}+\frac{H_2}{H^*_2}K_2{\rm e}^{\zeta_2+\zeta^*_2}
+ H_1 H_2 K_{12}{\rm e}^{\zeta_1+\zeta_2}+\frac{1}{H^*_1H^*_2} K^*_{12}{\rm e}^{\zeta^*_1+\zeta^*_2}
+ \frac{H_1}{H^*_2}\tilde{K}_{12}{\rm e}^{\zeta_1+\zeta^*_2}+ \frac{H_2}{H^*_1}\tilde{K}^*_{12}{\rm e}^{\zeta^*_1+\zeta_2}\ \ \ \\
&&\nonumber \ \ \ \ \ \
+\frac{H_1H_2}{H^*_1} K_1K_{12}\tilde{K}^*_{12} {\rm e}^{\zeta_1+\zeta^*_1+\zeta_2}
+\frac{H_1}{H^*_1H^*_2} K_1K^*_{12}\tilde{K}_{12} {\rm e}^{\zeta_1+\zeta^*_1+\zeta^*_2}\\
&&\nonumber \ \ \ \ \ \
+\frac{H_1H_2}{H^*_2} K_2K_{12}\tilde{K}_{12}{\rm e}^{\zeta_1+\zeta_2+\zeta^*_2}
+\frac{H_2}{H^*_1H^*_2} K_2K^*_{12}\tilde{K}^*_{12}{\rm e}^{\zeta^*_1+\zeta_2+\zeta^*_2}\\
&& \ \ \ \ \ \
+\frac{H_1H_2}{H^*_1H^*_2} K_1K_2 K_{12}K^*_{12}\tilde{K}_{12}\tilde{K}^*_{12} {\rm e}^{\zeta_1+\zeta^*_1+\zeta_2+\zeta^*_2},
\end{eqnarray}
with
\begin{eqnarray*}
&& K_i=1-\frac{\omega^2_i}{(\Omega_i-\Omega^*_i)^2},\ \
 H_i=\frac{2\Omega_i+2a-\omega_i}{2\Omega_i+2a+\omega_i},\ \ \\
&& K_{12}=\frac{4(\Omega_1-\Omega_2)^2-(\omega_1-\omega_2)^2}{4(\Omega_1-\Omega_2)^2-(\omega_1+\omega_2)^2},\\
&&
\tilde{K}_{12}=\frac{4(\Omega_1-\Omega^*_2)^2-(\omega_1+\omega_2)^2}{4(\Omega_1-\Omega^*_2)^2-(\omega_1-\omega_2)^2},
\end{eqnarray*}
and
\begin{eqnarray*}
\zeta_i=-\textrm{i}w_i x
+ \left[ \frac{4{\rm i}\sigma \rho^2\omega_i}{(2\Omega_i+2a)^2-\omega^2_i} -2{\rm i}\omega_i\Omega_i \right]y
+ [-\frac{4{\rm i}\sigma \rho^2\omega_i}{(2\Omega_i+2a)^2-\omega^2_i}-2{\rm i}a\omega_i] t
+\zeta_{i,0},
\end{eqnarray*}
for $i=1,2$.
This solution contains the two general breather solution, the second homoclinic orbit and the mixed solution for the two-dimensional YO system, which are shown in Fig. \ref{two-gen-breather-2d}--\ref{two-breather-homoclinic-2d}.
In Fig. \ref{two-gen-breather-2d}, one can observe that two breather are two localized soliton moving on the constant background,
and undergo a similar elastic collision along with the time changes. Fig. \ref{two-homclinic-2d} illustrates the second homoclinic orbit, which is the combination homoclinic solution with two modes. The mixed solution consisting of one-breather and one homoclinic orbit is displayed in Fig. \ref{two-breather-homoclinic-2d}. In the mixed case, although the breather always exist and the homoclinic orbit is localized in time, the maximum amplitude of the homoclinic orbit decreases obviously in the intermediate time.

\subsection{Rational solution}
In order to obtain the rational solutions for the YO system, we follow the procedure of Ablowitz and Satsuma \cite{ablowitz1978solitons,satsuma1979two}.
Starting from the singular soliton (${\rm e}^{\zeta_{k,0}}=-1$), and taking the long wave limit ($\omega_k \rightarrow 0$),
one find the rational solution as follows:

\textbf{Theorem 3.3} The rational solutions for two-dimensional YO system are
\begin{eqnarray}
 S=\rho{\rm e}^{-{\rm i}\alpha t}\frac{g}{f},\ \ L=\alpha+2\left( \ln f \right)_{xx},
\end{eqnarray}
where $f= |F_{k,l}|$, $g= |G_{k,l}|$ and the matrix elements are defined by
\begin{eqnarray*}
&& F_{k,k}= \left( \begin {array}{cc}
\theta_k &  - \frac{1}{\Omega_k-\Omega^*_k} \\
- \frac{1}{\Omega_k-\Omega^*_k}  &  \theta^*_k
 \end {array} \right),\\
&& G_{k,k}= \left( \begin {array}{cc}
\theta_k-\frac{1}{\Omega_k+a} &
  - \frac{1}{\Omega_k-\Omega^*_k}\frac{\Omega_k+a}{\Omega^*_k+a} \\
- \frac{1}{\Omega_k-\Omega^*_k} \frac{\Omega^*_k+a}{\Omega_k+a}  &
 \theta^*_k+\frac{1}{\Omega^*_k+a}
 \end {array} \right),\\
&& F_{k,l}=\left( \begin {array}{cc}
\frac{1}{-(\Omega_k-\Omega_l)} &
\frac{1}{-(\Omega_k-\Omega^*_l)}\\
\frac{1}{(\Omega^*_k-\Omega_l)} &
\frac{1}{(\Omega^*_k-\Omega_l)}
 \end {array} \right),\\
&& G_{k,l}=\left( \begin {array}{cc}
\frac{1}{-(\Omega_k-\Omega_l)} \frac{\Omega_k+a  }{\Omega_l+a  }  &
\frac{1}{-(\Omega_k-\Omega^*_l)} \frac{\Omega_k+a  }{\Omega^*_l+a } \\
\frac{1}{(\Omega^*_k-\Omega_l)} \frac{\Omega^*_k+a  }{\Omega_l+a } &
\frac{1}{(\Omega^*_k-\Omega_l)} \frac{\Omega^*_k+a  }{\Omega^*_l+a }
 \end {array} \right),
\end{eqnarray*}
with
\begin{eqnarray*}
&& \theta_k=-{\rm i}x + \left[\frac{{\rm i}\sigma\rho^2}{(\Omega_k+a)^2} -2{\rm i}\Omega_k \right]y - \left[\frac{{\rm i}\sigma\rho^2}{(\Omega_k+a)^2} +2{\rm i}a \right]t,
\end{eqnarray*}
where $\Omega_k$ are complex parameters for $k=1,2,\cdots,M$.

For example, when $M=1$, the simplest rational solution is given by
\begin{eqnarray}
&& f=|\theta_1|^2-\frac{1}{(\Omega_1-\Omega^*_1)^2},\ \
 g=\vartheta_1\vartheta'_1-\frac{1}{(\Omega_1-\Omega^*_1)^2},
\end{eqnarray}
with
\begin{eqnarray*}
&& \theta_1=-{\rm i}x + \left[\frac{{\rm i}\sigma\rho^2}{(\Omega_1+a)^2} -2{\rm i}\Omega_1 \right]y - \left[\frac{{\rm i}\sigma\rho^2}{(\Omega_1+a)^2} +2{\rm i}a \right]t,\\
&& \vartheta_1=\theta_1-\frac{1}{\Omega_1+a}, \ \ \vartheta'_1=\theta^*_1+\frac{1}{\Omega^*_1+a},
\end{eqnarray*}

Let $\Omega_1=\Omega_{1R}+{\rm i}\Omega_{1I}$, the rational solution can be rewritten as
\begin{eqnarray}\label{byo-58}
&&f=\theta_1\theta^*_1 + \theta_0,\ \ g=(\theta_1+b_1+{\rm i}b_2)(\theta^*_1-b_1+{\rm i}b_2)+\theta_0,
\end{eqnarray}
with $\theta_1=-{\rm i}x +(c_1+{\rm i}c_2)y + (d_1+{\rm i}d_2)t$ and
\begin{eqnarray*}
&&\theta_0=\frac{1}{4\Omega^2_{1I}},\ \
 c_1= -d_1 +2\Omega_{1I},\ \
 c_2= -d_2-2a -2\Omega_{1R},\\
&&b_1=-\frac{\Omega_{1R}+a}{(\Omega_{1R}+a)^2+\Omega^2_{1I}},\ \
b_2=\frac{\Omega_{1I}}{(\Omega_{1R}+a)^2+\Omega^2_{1I}},\\
&& d_1=\frac{2\sigma\rho^2\Omega_{1I}(\Omega_{1R}+a)}{[(\Omega_{1R}+a)-\Omega^2_{1I}]^2+4\Omega^2_{1I}(\Omega_{1R}+a)^2},\\
&& d_2=-\frac{\sigma\rho^2[(\Omega_{1R}+a)-\Omega^2_{1I}]}{[(\Omega_{1R}+a)-\Omega^2_{1I}]^2+4\Omega^2_{1I}(\Omega_{1R}+a)^2}-2a.
\end{eqnarray*}
Then the last expression of the rational solution reads
\begin{eqnarray}
\label{byo-59}&& S=\rho{\rm e}^{-{\rm i}\alpha t}\left[1- \frac{ 2{\rm i}b_1(-x+c_2y+d_2t) -2{\rm i}b_2 (c_1y+d_1t) +b^2_1+b^2_2}{(-x+c_2y+d_2t)^2+(c_1y+d_1t)^2+\theta_0} \right],\\
\label{byo-60}&& L=\alpha- 4\frac{(-x+c_2y+d_2t)^2- (c_1y+d_1t)^2-\theta_0}{[(-x+c_2y+d_2t)^2+(c_1y+d_1t)^2+\theta_0]^2 },
\end{eqnarray}

\begin{figure}[!htbp]
\centering
{\includegraphics[height=1.2in,width=3.2in]{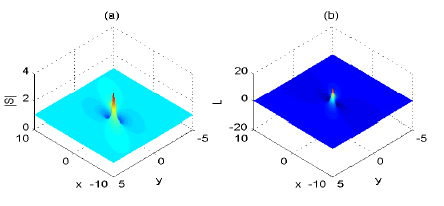}}
\caption{  One-lump for two-dimensional YO system defined by Eqs. (\ref{byo-59})--(\ref{byo-60}) with the parameters $\sigma=\rho=1$, $a=0$,  $\Omega_1=\frac{1}{4}+\frac{4}{5}{\rm i}$ at the time $t=0$. \label{one-lump-2d}}
\end{figure}

(i) Lump solution. From (\ref{byo-59})-(\ref{byo-60}), one can see that $(S,L)$ are constants along the $[x(t),y(t)]$ trajectory where
\begin{eqnarray*}
-x+c_2y=-d_2t,\ \ c_1y=-d_1t.
\end{eqnarray*}
Meanwhile, at any fixed time, when $(x,y)$ goes to infinity, $(S,L)\rightarrow (\rho{\rm e}^{-{\rm i}\alpha t},\alpha)$.
Hence we have a permanent lump moving on a constant background. One example of one-lump profiles is shown in Fig. \ref{one-lump-2d} at the given time $t=0$.

(ii) Rogue wave.  If $c_1=0$, ie.,
\begin{eqnarray}\label{byo-61}
-\frac{2\sigma\rho^2\Omega_{1I}(\Omega_{1R}+a)}{[(\Omega_{1R}+a)-\Omega^2_{1I}]^2+4\Omega^2_{1I}(\Omega_{1R}+a)^2}+2\Omega_{1I}=0,
\end{eqnarray}
the solution is a line wave, but it is
not amoving line soliton. As $t\rightarrow \pm \infty$, this line wave
goes to a uniform constant background; in the intermediate
times, it rises to a higher amplitude.

Specifically,  this solution describes a line rogue wave with the line
oriented in the $(c_2,1)$ direction of the $(x,y)$ plane, thus
the fundamental rogue waves in the two-dimensional YO system are line
rogue waves. The orientation angle $\beta$ of this line rogue wave
is $\beta=\tan^{-1}(c_2)$, and its width is inversely proportional
to $\sqrt{1+c^2_2}$.

\begin{figure*}[!htbp]
\centering
{\includegraphics[height=2.8in,width=6.5in]{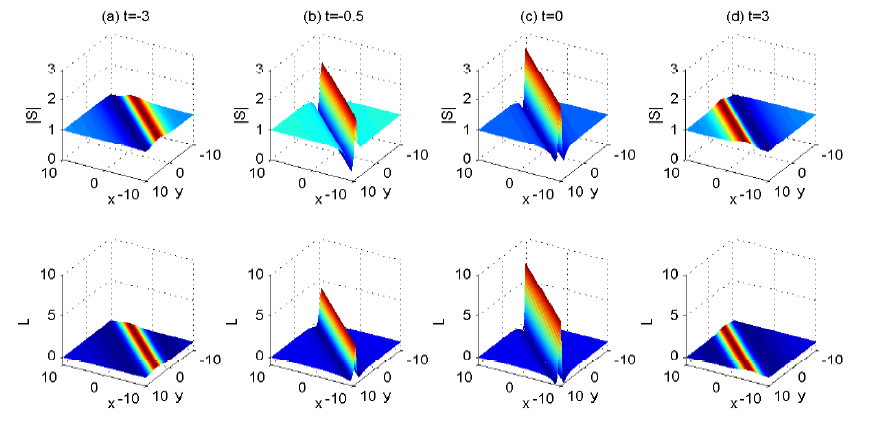}}
\caption{  One-rogue wave for the two-dimensional YO system defined by Eqs. (\ref{byo-59})--(\ref{byo-60}) with the parameters $\sigma=\rho=1$, $a=0$,  $\Omega_1=\frac{1}{4}+\frac{4}{5}{\rm i}$ at the time $t=0$. \label{one-rogue-2d}}
\end{figure*}

As $t \rightarrow \pm \infty$, the solution $(S,L)$ uniformly
approach the constant background $\rho{\rm e}^{-{\rm i}\alpha t}$ and $\alpha$ everywhere in the
$(x,y)$ plane; but in the intermediate times, $|S|$ and $L$ reach
maximum amplitude $|\frac{\rho}{\Omega_1+a}|$ and $\alpha+4\Omega^2_{1I}$ respectively at the center ($-x+c_2y=0$) of the line wave at
time $t=0$. One-rogue wave for the two-dimensional YO system is illustrated in Fig. \ref{one-rogue-2d}.

In Ref.\cite{ohta2012rogue,ohta2013dynamics}, Ohta \emph{et al.} have shown that the fundamental rogue waves in the DS equations are
two-dimensional counterparts of the fundamental (Peregrine) rogue waves in the NLS equation.
Very similarly, for the YO system, the fundamental rogue waves of the two-dimensional case are viewed as the fundamental ones of the one-dimensional case.
Here, if we further take $c_2=0$, ie., $\frac{\sigma\rho^2[(\Omega_{1R}+a)-\Omega^2_{1I}]}{[(\Omega_{1R}+a)-\Omega^2_{1I}]^2+4\Omega^2_{1I}(\Omega_{1R}+a)^2}-2\Omega_{1R}=0$ in Eq.(\ref{byo-58}), the solution is independent of $y$. Thus, the two-dimensional YO system is reduced to the one-dimensional case, and the this fundamental rogue wave of the two-dimensional YO system is reduced to the fundamental rogue wave of the one-dimensional YO system.

Assuming $\Omega_k=\Omega_{kR}+{\rm i}\Omega_{kI}$ for $k=1,2,\cdots,M$, multi-rogue wave for the two-dimensional YO system can be obtained in theorem 3.1 by restricting the parameters' constraints conditions
\begin{eqnarray}\label{byo-62}
-\frac{2\sigma\rho^2\Omega_{kI}(\Omega_{kR}+a)}{[(\Omega_{kR}+a)-\Omega^2_{kI}]^2+4\Omega^2_{kI}(\Omega_{kR}+a)^2}+2\Omega_{kI}=0.
\end{eqnarray}
These rogue wave solutions describe the interaction of $M$ individual fundamental rogue waves.

When $M=2$, the solution in theorem 3.1 can be written as
\begin{eqnarray}
\label{byo-63}&&  S=\rho{\rm e}^{-{\rm i}\alpha t}\frac{g}{f},\ \ L=\alpha+2\left( \ln f \right)_{xx},\\
\label{byo-64}&&\nonumber f=|\theta_1|^2|\theta_2|^2-B^2_{34}|\theta_1|^2-B^2_{12}|\theta_2|^2
+B^2_{24}\theta_1\theta_2+B^2_{13}\theta^*_1\theta^*_2
-B^2_{23}\theta_1\theta^*_2-B^2_{14}\theta^*_1\theta_2\\
&&\ \ \ \ \ \ +(B_{12}B_{34}-B_{14}B_{23}-B_{13}B_{24})^2,\\
\label{byo-65}&&\nonumber g=\vartheta_1\vartheta_2\vartheta'_1\vartheta'_2
-B^2_{34}\vartheta_1\vartheta'_1-B^2_{12}\vartheta_2\vartheta'_2
+B^2_{24}\vartheta_1\vartheta_2+B^2_{13}\vartheta'_1\vartheta'_2
-B^2_{23}\vartheta_1\vartheta'_2-B^2_{14}\vartheta'_1\vartheta_2\\
&&\ \ \ \ \ \ +(B_{12}B_{34}-B_{14}B_{23}-B_{13}B_{24})^2,
\end{eqnarray}
with
\begin{eqnarray*}
&& \theta_k=-{\rm i}x + \left[\frac{{\rm i}\sigma\rho^2}{(\Omega_k+a)^2} -2{\rm i}\Omega_k \right]y - \left[\frac{{\rm i}\sigma\rho^2}{(\Omega_k+a)^2} +2{\rm i}a \right]t,\\
&& \vartheta_k=\theta_k-\frac{1}{\Omega_k+a}, \ \ \vartheta'_k=\theta^*_k+\frac{1}{\Omega^*_k+a},
\end{eqnarray*}
where
\begin{eqnarray*}
&& B_{12}=\frac{1}{\Omega^*_1-\Omega_1},\ \ B_{13}=B^*_{24}=\frac{1}{\Omega_1-\Omega_2},\ \
B_{14}=-B^*_{23}=\frac{1}{\Omega^*_2-\Omega_1},\ \ B_{34}=\frac{1}{\Omega^*_2-\Omega_2}.
\end{eqnarray*}

\begin{figure*}[!htbp]
\centering
{\includegraphics[height=2.1in,width=4.3in]{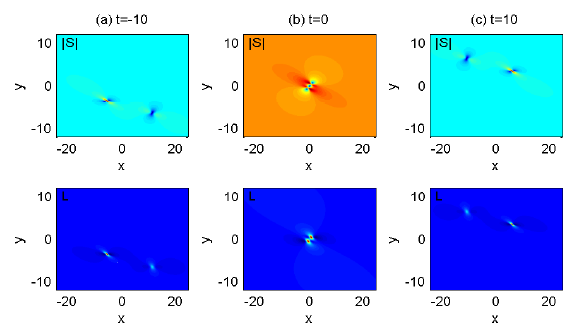}}
\caption{  The two-lump for two-dimensional YO system defined by Eqs. (\ref{byo-63})--(\ref{byo-65}) with the parameters $\sigma=\rho=1$, $a=0$,  $\Omega_1=\frac{1}{4}+\frac{4}{5}{\rm i}$ and $\Omega_2=\frac{2}{3}+\frac{2}{5}{\rm i}$. \label{two-lump-2d} }
\end{figure*}

\begin{figure*}[!htbp]
\centering
{\includegraphics[height=2.1in,width=6in]{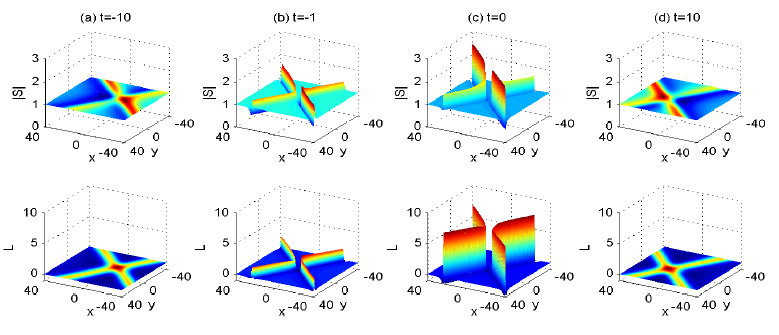}}
\caption{  The two-rogue wave for two-dimensional YO system defined by Eqs. (\ref{byo-63})--(\ref{byo-65}) and $(\ref{byo-62})(k=1,2)$ with the parameters $\sigma=\rho=1$, $a=0$,  $\Omega_1=-\frac{1}{4}+\frac{\sqrt{7}}{4}{\rm i}$ and $\Omega_2=-\frac{1}{2}+\frac{\sqrt{2\sqrt{2}-1}}{2}{\rm i}$. \label{two-rogue-2d} }
\end{figure*}

\begin{figure*}[!htbp]
\centering
{\includegraphics[height=2.1in,width=6in]{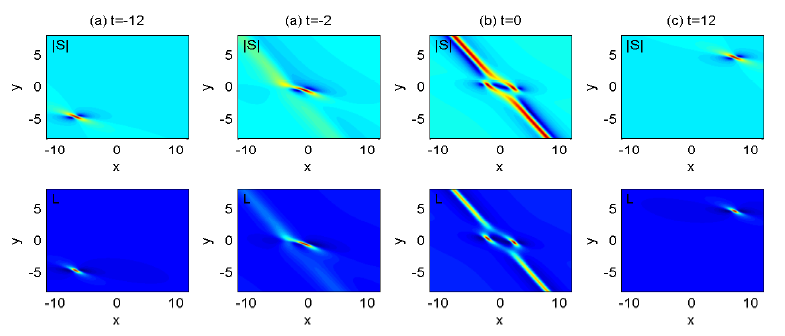}}
\caption{  The mixed solution consisting of one lump and one rogue wave for two-dimensional YO system defined by Eqs. (\ref{byo-63})--(\ref{byo-65}) and $(\ref{byo-62})(k=1)$ with the parameters $\sigma=\rho=1$, $a=0$,  $\Omega_1=-\frac{1}{4}+\frac{\sqrt{7}}{4}{\rm i}$ and $\Omega_2=\frac{1}{4}+\frac{3}{4}{\rm i}$. \label{two-lumprogue-2d} }
\end{figure*}

This solution contains two-lump solution, two-rogue wave and the mixed solution consisting of one-lump and one-rogue wave for the two-dimensional YO system.
The two-lump is shown in Fig. \ref{two-lump-2d}, in which two lumps undergo a similar elastic collision.
This phenomenon is same to the two-general breather in the previous section: when the period increases to infinity, two-general breather solution becomes the two-lump solution.
In Fig. \ref{two-rogue-2d}, we illustrate two-rogue wave of the two-dimensional YO system.
It is seen that two line rogue waves arise from the constant background, then reach their maximum amplitude in the intermediate times and finally disappear into the background again.
Besides, two line rogue waves have intersection with lower amplitude first. When they reach their maximum amplitude in the intermediate times, the wave pattern forms two curvy wave fronts which are completely separated. Then two line rogue waves have intersection with lower amplitude again.
The mixed solution consisting of one-lump and one-rogue wave is displayed in Fig. \ref{two-lumprogue-2d}.
Obviously, the lump always exist and move on the constant background. When the amplitude of the line rogue wave reach the maximum value, this line rogue wave cross over the lump directly.

\textbf{Remark 3.1} For the rational solution of the one-dimensional YO system, only one-rational solution can be obtained.
The reason is that, if the rational solution in theorem 3.1 is independent of $y$, the parameters constraint condition $\frac{{\rm i}\sigma\rho^2}{(\Omega_k+a)^2} -2{\rm i}\Omega_k=0$ need to be satisfied.
For the given parameters $\sigma, a$ and $\rho$, there only exist three solutions $\Omega_k$, in which two are complex conjugation and other is real.
However, from the expression in theorem 3.1, we know that $Im(\Omega_k)\neq0$ and $\Omega_i\neq\Omega^*_j$ ensure the solution nonsingular.
Consequently, we only get one-rational solution for the one-dimensional YO system through the further reduction.

\newpage
\subsection{Rational-exp solution}

Starting from the singular soliton (${\rm e}^{\xi_{k,0}+\eta_{k,0}}=-1$), $k_{s_1}<k<k_{s_{\tilde{N}}}$, and taking the long wave limit ($\omega_k \rightarrow 0$),
one find the rational-exp solution as follows:

\textbf{Theorem 3.4} The $\tilde{N}$-rational-$\tilde{N}'$-exp solutions for two-dimensional YO system are
\begin{eqnarray}
 S=\rho{\rm e}^{-{\rm i}\alpha t}\frac{g}{f},\ \ L=\alpha+2\left( \ln f \right)_{xx},
\end{eqnarray}
where $f= \Delta_0|F_{k,l}|$, $g= \Delta_0|G_{k,l}|$ and $\Delta_0= \textmd{e}^{\sum^{s'_{\tilde{N}'}}_{k=s'_1}\zeta_k+\zeta^*_k} \prod^{s'_{\tilde{N}'}}_{k=s'_1}\omega^2_k$, and the matrix elements are defined as follows:

\begin{eqnarray}
 |F_{k,l}|=\left|  \begin {array}{cc} A & B\\ C & D \end {array} \right|,\ \
|G_{k,l}|=\left|  \begin {array}{cc} \mathcal{A} & \mathcal{B}\\ \mathcal{C} & \mathcal{D} \end {array} \right|,
\end{eqnarray}
where $A$ and $\mathcal{A}$ are $2\tilde{N}\times 2\tilde{N}$ matrices defined by
\begin{eqnarray*}
&& A_{k,k}= \left( \begin {array}{cc}
\theta_k &  - \frac{1}{\Omega_k-\Omega^*_k} \\
- \frac{1}{\Omega_k-\Omega^*_k}  &  \theta^*_k
 \end {array} \right),\ \
\mathcal{A}_{k,k}= \left( \begin {array}{cc}
\theta_k-\frac{1}{\Omega_k+a} &
  - \frac{1}{\Omega_k-\Omega^*_k}\frac{\Omega_k+a}{\Omega^*_k+a} \\
- \frac{1}{\Omega_k-\Omega^*_k} \frac{\Omega^*_k+a}{\Omega_k+a}  &
 \theta^*_k+\frac{1}{\Omega^*_k+a}
 \end {array} \right),\\
&& A_{k,l}=\left( \begin {array}{cc}
\frac{1}{-\Omega_k+\Omega_l} &
\frac{1}{-\Omega_k+\Omega^*_l}\\
\frac{1}{\Omega^*_k-\Omega_l} &
\frac{1}{\Omega^*_k-\Omega_l}
 \end {array} \right),\ \
 \mathcal{A}_{k,l}=\left( \begin {array}{cc}
\frac{1}{-\Omega_k+\Omega_l} \frac{\Omega_k+a }{\Omega_l+a  }  &
\frac{1}{-\Omega_k+\Omega^*_l} \frac{\Omega_k+a  }{\Omega^*_l+a  } \\
\frac{1}{\Omega^*_k-\Omega_l} \frac{\Omega^*_k+a  }{\Omega_l+a  } &
\frac{1}{\Omega^*_k-\Omega_l} \frac{\Omega^*_k+a }{\Omega^*_l+a  }
 \end {array} \right),
\end{eqnarray*}
$D$ and $\mathcal{D}$ are $2\tilde{N}'\times 2\tilde{N}'$ matrices defined by
\begin{eqnarray*}
&& D_{k,k}= \left( \begin {array}{cc}
\frac{1}{\omega_k {\rm e}^{\zeta_k}} + \frac{1}{\omega_k} &  - \frac{1}{\Omega_k-\Omega^*_k} \\
- \frac{1}{\Omega_k-\Omega^*_k}  &  \frac{1}{\omega_k {\rm e}^{\zeta^*_k}} + \frac{1}{\omega_k}
 \end {array} \right),\\
&& \mathcal{D}_{k,k}= \left( \begin {array}{cc}
\frac{1}{\omega_k {\rm e}^{\zeta_k}} + \frac{1}{\omega_k}\frac{\Omega_k+a+ \frac{\omega_k}{2}}{\Omega_k+a- \frac{\omega_k}{2}} &
  - \frac{1}{\Omega_k-\Omega^*_k}\frac{\Omega_k+a- \frac{\omega_k}{2}}{\Omega^*_k+a- \frac{\omega_k}{2}} \\
- \frac{1}{\Omega_k-\Omega^*_k} \frac{\Omega^*_k+a+ \frac{\omega_k}{2}}{\Omega_k+a+ \frac{\omega_k}{2}}  &
 \frac{1}{\omega_k {\rm e}^{\zeta^*_k}} + \frac{1}{\omega_k}
 \frac{\Omega^*_k+a+ \frac{\omega_k}{2}}{\Omega^*_k+a- \frac{\omega_k}{2}}
 \end {array} \right),\\
&& D_{k,l}=\left( \begin {array}{cc}
\frac{1}{-(\Omega_k-\Omega_l)+\frac{\omega_k+ \omega_l}{2}} &
\frac{1}{-(\Omega_k-\Omega^*_l)+\frac{\omega_k- \omega_l}{2}}\\
\frac{1}{(\Omega^*_k-\Omega_l)+\frac{\omega_k- \omega_l}{2}} &
\frac{1}{(\Omega^*_k-\Omega_l)+\frac{\omega_k+ \omega_l}{2}}
 \end {array} \right),\\
&& \mathcal{D}_{k,l}=\left( \begin {array}{cc}
\frac{1}{-(\Omega_k-\Omega_l)+\frac{ \omega_k+ \omega_l}{2}} \frac{\Omega_k+a - \frac{\omega_k}{2} }{\Omega_l+a + \frac{\omega_l}{2} }  &
\frac{1}{-(\Omega_k-\Omega^*_l)+\frac{\omega_k- \omega_l}{2}} \frac{\Omega_k+a -\frac{\omega_k}{2} }{\Omega^*_l+a -  \frac{\omega_l}{2} } \\
\frac{1}{(\Omega^*_k-\Omega_l)+\frac{ \omega_k- \omega_l}{2}} \frac{\Omega^*_k+a + \frac{\omega_k}{2} }{\Omega_l+a + \frac{\omega_l}{2} } &
\frac{1}{(\Omega^*_k-\Omega_l)+\frac{ \omega_k+\omega_l}{2}} \frac{\Omega^*_k+a + \frac{\omega_k}{2} }{\Omega^*_l+a -\frac{\omega_l}{2} }
 \end {array} \right),
\end{eqnarray*}
$B$ and $\mathcal{B}$ are $2\tilde{N}\times 2\tilde{N}'$ matrices defined by
\begin{eqnarray*}
&& B_{k,l}=\left( \begin {array}{cc}
\frac{1}{-(\Omega_k-\Omega_l)+\frac{\omega_l}{2}} &
\frac{1}{-(\Omega_k-\Omega^*_l)+\frac{\omega_l}{2}}\\
\frac{1}{(\Omega^*_k-\Omega_l)+\frac{\omega_l}{2}} &
\frac{1}{(\Omega^*_k-\Omega_l)+\frac{\omega_l}{2}}
 \end {array} \right),\\
&& \mathcal{B}_{k,l}=\left( \begin {array}{cc}
\frac{1}{-(\Omega_k-\Omega_l)+\frac{ \omega_l}{2}} \frac{\Omega_k+a }{\Omega_l+a +\frac{\omega_l}{2} }  &
\frac{1}{-(\Omega_k-\Omega^*_l)+\frac{ \omega_l}{2}} \frac{\Omega_k+a  }{\Omega^*_l+a -  \frac{\omega_l}{2} } \\
\frac{1}{(\Omega^*_k-\Omega_l)+\frac{ \omega_l}{2}} \frac{\Omega^*_k+a  }{\Omega_l+a + \frac{\omega_l}{2} } &
\frac{1}{(\Omega^*_k-\Omega_l)+\frac{\omega_l}{2}} \frac{\Omega^*_k+a  }{\Omega^*_l+a - \frac{\omega_l}{2} }
 \end {array} \right),
\end{eqnarray*}
$C$ and $\mathcal{C}$ are $2\tilde{N}'\times 2\tilde{N}$ matrices defined by
\begin{eqnarray*}
&& C_{k,l}=\left( \begin {array}{cc}
\frac{1}{-(\Omega_k-\Omega_l)+\frac{\omega_k}{2}} &
\frac{1}{-(\Omega_k-\Omega^*_l)+\frac{\omega_k}{2}}\\
\frac{1}{(\Omega^*_k-\Omega_l)+\frac{\omega_k}{2}} &
\frac{1}{(\Omega^*_k-\Omega_l)+\frac{\omega_k}{2}}
 \end {array} \right),\\
&& \mathcal{C}_{k,l}=\left( \begin {array}{cc}
\frac{1}{-(\Omega_k-\Omega_l)+\frac{\omega_k}{2}} \frac{\Omega_k+a -\frac{\omega_k}{2} }{\Omega_l+a  }  &
\frac{1}{-(\Omega_k-\Omega^*_l)+\frac{\omega_k}{2}} \frac{\Omega_k+a -\frac{\omega_k}{2} }{\Omega^*_l+a  } \\
\frac{1}{(\Omega^*_k-\Omega_l)+\frac{\omega_k}{2}} \frac{\Omega^*_k+a +\frac{\omega_k}{2} }{\Omega_l+a  } &
\frac{1}{(\Omega^*_k-\Omega_l)+\frac{\omega_k}{2}} \frac{\Omega^*_k+a +\frac{\omega_k}{2} }{\Omega^*_l+a  }
 \end {array} \right),
\end{eqnarray*}
and
\begin{eqnarray*}
&& \theta_k=-{\rm i}x + \left[\frac{{\rm i}\sigma\rho^2}{(\Omega_k+a)^2} -2{\rm i}\Omega_k \right]y - \left[\frac{{\rm i}\sigma\rho^2}{(\Omega_k+a)^2} +2{\rm i}a \right]t,\\
&& \zeta_{k} = -\textrm{i}\omega_k x
+ \left[ \frac{4{\rm i}\sigma \rho^2\omega_k}{(2\Omega_k+2a)^2-\omega^2_k} -2{\rm i}\omega_k\Omega_k \right]y
+ [-\frac{4{\rm i}\sigma \rho^2\omega_k}{(2\Omega_k+2a)^2-\omega^2_k}-2{\rm i}a\omega_k] t +\zeta_{k,0}.
\end{eqnarray*}

When two-dimensional YO system reduces to one-dimensional case, one only obtain one-rational-$\tilde{N}'$-exp solution (one-rogue-wave-$\tilde{N}'$-breather solution) with the following constraints conditions:
\begin{eqnarray}
\label{byo-67}&& \frac{\sigma\rho^2}{(\Omega_1+a)^2} -2\Omega_1=0,\\
\label{byo-68} && \frac{2\sigma \rho^2}{(2\Omega_k+2a)^2-\omega^2_k} -\Omega_k =0,
\end{eqnarray}
for $k=1,2,\cdots,\tilde{N}'$.

\begin{figure*}[!htbp]
\centering
{\includegraphics[height=2.1in,width=6in]{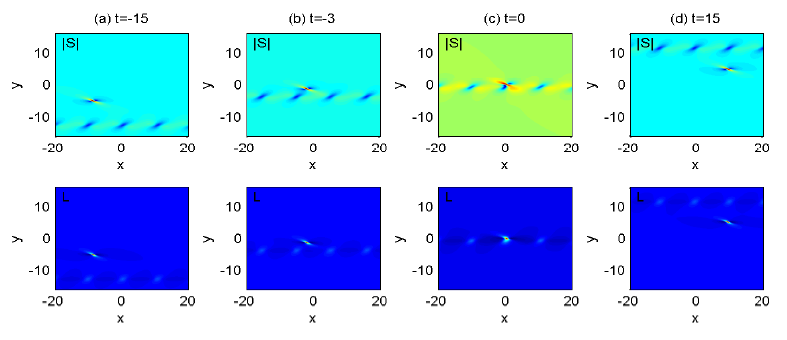}}
\caption{  The mixed solution consisting of one lump and one general breather for two-dimensional YO system defined by Eqs. (\ref{byo-69})--(\ref{byo-71}) with the parameters $\sigma=\rho=1$, $a=\zeta_{2,0}=0$,  $\Omega_1=\frac{1}{4}+\frac{4}{5}{\rm i}$, $\Omega_2=\frac{2}{3}+\frac{1}{5}{\rm i}$ and $\omega_2=\frac{3}{5}$. \label{two-lump-breather-2d} }
\end{figure*}

\begin{figure*}[!htbp]
\centering
{\includegraphics[height=2.1in,width=6in]{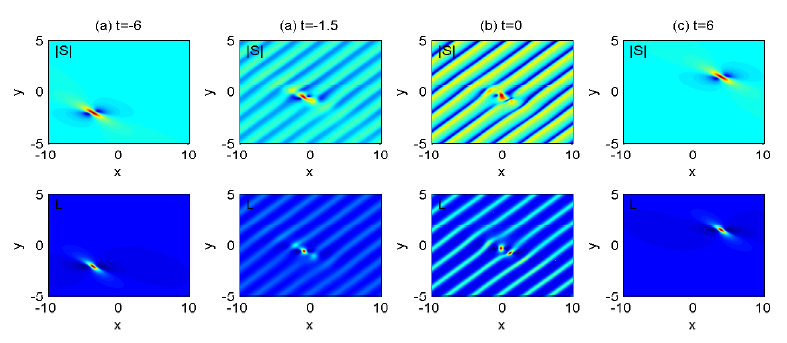}}
\caption{  The mixed solution consisting of one lump and one homoclinic orbit for two-dimensional YO system defined by Eqs. (\ref{byo-69})--(\ref{byo-71}) and (\ref{byo-51}) with the parameters $\sigma=\rho=1$, $a=\zeta_{2,0}=0$,  $\Omega_1=\frac{1}{4}+\frac{4}{5}{\rm i}$, $\Omega_2=-1+\frac{1}{2}{\rm i}$ and $\omega_2=\sqrt{3}$. \label{two-lump-homoclinic-2d} }
\end{figure*}

\begin{figure*}[!htbp]
\centering
{\includegraphics[height=2.1in,width=6in]{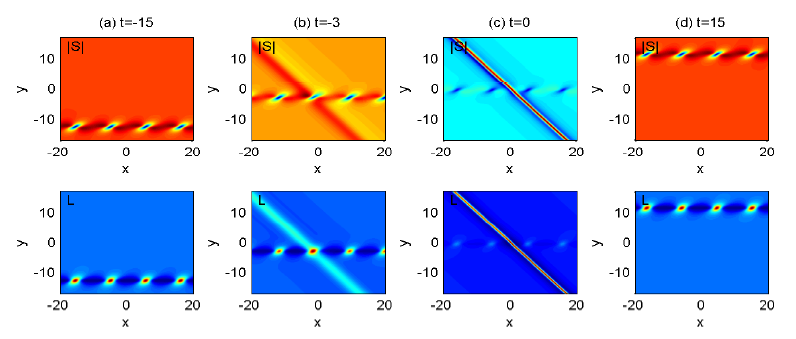}}
\caption{  The mixed solution consisting of the breather and one rogue wave for two-dimensional YO system defined by Eqs. (\ref{byo-69})--(\ref{byo-71}) and (\ref{byo-61}) with the parameters $\sigma=\rho=1$, $a=\zeta_{2,0}=0$,  $\Omega_1=-\frac{1}{4}+\frac{\sqrt{7}}{4}{\rm i}$, $\Omega_2=\frac{2}{3}+\frac{1}{5}{\rm i}$ and $\omega_2=\frac{3}{5}$. \label{two-rogue-breather-2d} }
\end{figure*}

\begin{figure*}[!htbp]
\centering
{\includegraphics[height=2.1in,width=6in]{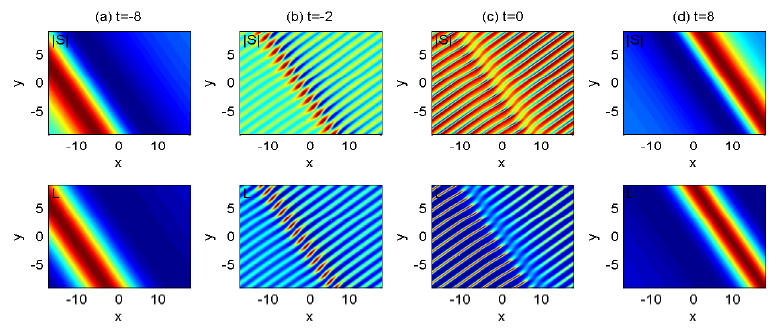}}
\caption{  The mixed solution consisting of one rogue wave and one homoclinic orbit for two-dimensional YO system defined by Eqs. (\ref{byo-69})--(\ref{byo-71}) and (\ref{byo-51}) and (\ref{byo-61}) with the parameters $\sigma=\rho=1$, $a=\zeta_{2,0}=0$,  $\Omega_1=-\frac{1}{4}+\frac{\sqrt{7}}{4}{\rm i}$, $\Omega_2=-1+\frac{1}{2}{\rm i}$ and $\omega_2=\sqrt{3}$. \label{two-rogue-homoclinic-2d} }
\end{figure*}

For example, when $\tilde{N}=\tilde{N}'=1$, one-rational-one-exp solution is given by
\begin{eqnarray}
\label{byo-69} && S=\rho{\rm e}^{-{\rm i}\alpha t}\frac{G}{F},\ \ L=\alpha+2\left( \ln \Delta_0 F \right)_{xx},\\
&& F=\left| \begin {array}{cccc} \theta_1& \frac{1}{\Omega^*_1-\Omega_1} &
\frac{1}{\Omega_2-\Omega_1 +\frac{\omega_2}{2}} & \frac{1}{\Omega^*_2-\Omega_1 -\frac{\omega_2}{2}} \\
\frac{1}{\Omega^*_1-\Omega_1} &\theta^*_1&
\frac{1}{\Omega^*_1-\Omega_2 -\frac{\omega_2}{2}} & \frac{1}{\Omega^*_1-\Omega^*_2 +\frac{\omega_2}{2}} \\
\frac{1}{\Omega_1-\Omega_2 +\frac{\omega_2}{2}} & \frac{1}{\Omega^*_1-\Omega_2 +\frac{\omega_2}{2}} & \frac{1}{ \omega_2 {\rm e}^{\zeta_2} } +\frac{1}{\omega_2} &  \frac{1}{\Omega^*_2-\Omega_2} \\
\frac{1}{\Omega^*_2-\Omega_1 +\frac{\omega_2}{2}} &\frac{1}{\Omega^*_2-\Omega^*_1 +\frac{\omega_2}{2}} & \frac{1}{\Omega^*_2-\Omega_2} & \frac{1}{ \omega_2 {\rm e}^{\zeta^*_2} } +\frac{1}{\omega_2}  \end {array} \right|,\\
\label{byo-71} && G=\left| \begin {array}{cccc} \theta_1-\frac{1}{\Omega_1+a} &
\frac{1}{\Omega^*_1-\Omega_1} \frac{\Omega_1+a}{\Omega^*_1+a} &
\frac{1}{\Omega_2-\Omega_1 +\frac{\omega_2}{2}} \frac{\Omega_1+a}{\Omega_2+a+\frac{\omega_2}{2}} &
 \frac{1}{\Omega^*_2-\Omega_1 -\frac{\omega_2}{2}}\frac{\Omega_1+a}{\Omega^*_2+a-\frac{\omega_2}{2}} \\
\frac{1}{\Omega^*_1-\Omega_1} \frac{\Omega^*_1+a}{\Omega_1+a} &
\theta^*_1 +\frac{1}{\Omega^*_1+a} &
\frac{1}{\Omega^*_1-\Omega_2 -\frac{\omega_2}{2}} \frac{\Omega^*_1+a}{\Omega_2+a+\frac{\omega_2}{2}} &
\frac{1}{\Omega^*_1-\Omega^*_2 +\frac{\omega_2}{2}} \frac{\Omega^*_1+a}{\Omega^*_2+a-\frac{\omega_2}{2}} \\
\frac{1}{\Omega_1-\Omega_2 +\frac{\omega_2}{2}} \frac{\Omega_2+a-\frac{\omega_2}{2}}{\Omega_1+a} &
\frac{1}{\Omega^*_1-\Omega_2 +\frac{\omega_2}{2}} \frac{\Omega_2+a-\frac{\omega_2}{2}}{\Omega^*_1+a} &
\frac{1}{ \omega_2 {\rm e}^{\zeta_2} } +\frac{1}{\omega_2}\frac{\Omega_2+a-\frac{\omega_2}{2}}{\Omega_2+a+\frac{\omega_2}{2}} &  \frac{1}{\Omega^*_2-\Omega_2}\frac{\Omega_2+a-\frac{\omega_2}{2}}{\Omega^*_2+a-\frac{\omega_2}{2}} \\
\frac{1}{\Omega^*_2-\Omega_1 +\frac{\omega_2}{2}} \frac{\Omega^*_2+a+\frac{\omega_2}{2}}{\Omega_1+a} &
\frac{1}{\Omega^*_2-\Omega^*_1 +\frac{\omega_2}{2}} \frac{\Omega^*_2+a+\frac{\omega_2}{2}}{\Omega^*_1+a} & \frac{1}{\Omega^*_2-\Omega_2}\frac{\Omega^*_2+a+\frac{\omega_2}{2}}{\Omega_2+a+\frac{\omega_2}{2}} &
\frac{1}{ \omega_2 {\rm e}^{\zeta^*_2} } +\frac{1}{\omega_2}\frac{\Omega^*_2+a+\frac{\omega_2}{2}}{\Omega^*_2+a-\frac{\omega_2}{2}}  \end {array} \right|,
\end{eqnarray}
where $\Delta_0=\omega^2_2  {\rm e}^{\zeta_2+\zeta^*_2}$ and
\begin{eqnarray*}
&& \theta_1=-{\rm i}x + \left[\frac{{\rm i}\sigma\rho^2}{(\Omega_1+a)^2} -2{\rm i}\Omega_1 \right]y - \left[\frac{{\rm i}\sigma\rho^2}{(\Omega_1+a)^2} +2{\rm i}a \right]t,\\
&& \zeta_{2} = -\textrm{i}\omega_2 x
+ \left[ \frac{4{\rm i}\sigma \rho^2\omega_2}{(2\Omega_2+2a)^2-\omega^2_2} -2{\rm i}\omega_2\Omega_2 \right]y
+ [-\frac{4{\rm i}\sigma \rho^2\omega_2}{(2\Omega_2+2a)^2-\omega^2_2}-2{\rm i}a\omega_2] t +\zeta_{2,0}.
\end{eqnarray*}
This solution represents the mixed solution consisting of  one lump (and one rogue wave) and one general breather (and one homoclinic orbit).
Examples of these mixed solutions are shown in Fig.\ref{two-lump-breather-2d}--\ref{two-rogue-homoclinic-2d}.

\begin{figure*}[!htbp]
\centering
{\includegraphics[height=1.2in,width=3.2in]{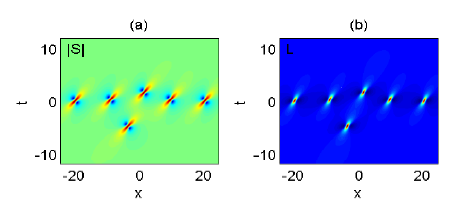}}
\caption{  The mixed solution consisting of one rogue wave and one general breather for one-dimensional YO system defined by Eqs. (\ref{byo-69})--(\ref{byo-71}) and (\ref{byo-67})-(\ref{byo-68})$(k=1)$ with the parameters $\sigma=\rho=1$, $a=\zeta_{2,0}=0$,  $\Omega_1=-\frac{2^{\frac{2}{3}}}{4} (1-\sqrt{3}{\rm i})$ and $\omega_2=\frac{3}{5}$. \label{one-rogue-one-breather-1d} }
\end{figure*}

Here, we also present one example of the mixed solution consisting of one rogue wave and one general breather for one-dimensional YO system as in Fig.\ref{one-rogue-one-breather-1d}.

\newpage

\section{Breather II}

\subsection{Complex conjugacy}
By taking $p_{2k-1}\rightarrow{\rm i}p_{2k-1}$, $p_{2k}\rightarrow-{\rm i}p_{2k}$, $q_{2k-1}\rightarrow{\rm i}q_{2k-1}$ and $q_{2k}\rightarrow-{\rm i}q_{2k}$ in Section 3.1.1, i.e.
\begin{eqnarray}
&& p_{2k-1}=-{\rm i}\Omega_k+{\rm i}\frac{\omega_k}{2},\ \ p_{2k}={\rm i}\Omega^*_k+{\rm i}\frac{\omega_k}{2},\ \ q_{2k-1}={\rm i}\Omega_k+{\rm i}\frac{\omega_k}{2},\ \ q_{2k}=-{\rm i}\Omega^*_k+{\rm i}\frac{\omega_k}{2},\ \
\end{eqnarray}
and $\xi_{2k-1,0}=\xi_{2k,0} \equiv \xi_{k,0},\ \ \eta_{2k-1,0}=\eta_{2k,0} \equiv \eta_{k,0}$, where $\Omega_k$ are complex parameters and $\omega_k,\xi_{k,0},\eta_{k,0}$ are real parameters for $k=1,2,\cdots,M$,
one has
\begin{eqnarray}
\nonumber && \xi_{2k-1}+\eta_{2k-1}
=\xi^*_{2k}+\eta^*_{2k} \\
&&\hspace{2.2cm} = \omega_k x
+ \left[ -\frac{4\sigma \rho^2\omega_k}{(2{\rm i}\Omega_k+2a)^2+\omega^2_k} +2{\rm i}\omega_k\Omega_k \right]y
+ [\frac{4\sigma \rho^2\omega_k}{(2{\rm i}\Omega_k+2a)^2+\omega^2_k}+2a\omega_k] t
+\xi_{k,0}+\eta_{k,0}.
\end{eqnarray}

\textbf{Theorem 4.1}
The breather solutions for the two-dimensional YO system are
\begin{eqnarray}\label{br2-01}
 S=\rho{\rm e}^{-{\rm i}\alpha t}\frac{g}{f},\ \ L=\alpha+2\left( \ln f \right)_{xx},
\end{eqnarray}
where $f= \Delta_0|F_{k,l}|$, $g= \Delta_0|G_{k,l}|$ and
$\Delta_0=\textmd{e}^{\sum^{M}_{i=1}\zeta_i+\zeta^*_i} \prod^{M}_{k=1}\omega^2_k$,
and the matrix elements are defined by
\begin{eqnarray*}
&& F_{k,k}= \left( \begin {array}{cc}
-\frac{{\rm i}}{\omega_k {\rm e}^{\zeta_k}} - \frac{{\rm i}}{\omega_k} &
\frac{{\rm i} }{\Omega_k+\Omega^*_k-\omega_k} \\
 \frac{ {\rm i} }{\Omega_k+\Omega^*_k+\omega_k}  &
 \frac{{\rm i}}{\omega_k {\rm e}^{\zeta^*_k}} + \frac{{\rm i}}{\omega_k}
 \end {array} \right),\\
&& G_{k,k}= \left( \begin {array}{cc}
-\frac{{\rm i}}{\omega_k {\rm e}^{\zeta_k}} - \frac{{\rm i}}{\omega_k}\frac{\Omega_k-{\rm i}a -\frac{\omega_k}{2}}{\Omega_k-{\rm i}a +\frac{\omega_k}{2}} &
\frac{ {\rm i} }{\Omega_k+\Omega^*_k-\omega_k} \frac{\Omega_k-{\rm i}a -\frac{\omega_k}{2} }{ -\Omega^*_k-{\rm i}a +\frac{\omega_k}{2} }\\
 \frac{{\rm i} }{\Omega_k+\Omega^*_k+\omega_k} \frac{\Omega^*_k+{\rm i}a + \frac{\omega_k}{2}}{-\Omega_k+{\rm i}a - \frac{\omega_k}{2}} &
 \frac{{\rm i}}{\omega_k {\rm e}^{\zeta^*_k}} + \frac{{\rm i}}{\omega_k} \frac{\Omega^*_k+{\rm i}a +\frac{\omega_k}{2}}{\Omega^*_k+{\rm i}a -\frac{\omega_k}{2} }
 \end {array} \right),\\
&& F_{k,l}=\left( \begin {array}{cc}
\frac{{\rm i}}{(\Omega_k-\Omega_l)-\frac{\omega_k+\omega_l}{2}} &
\frac{{\rm i}}{(\Omega_k+\Omega^*_l)-\frac{\omega_k+\omega_l}{2}}\\
\frac{{\rm i}}{(\Omega^*_k+\Omega_l)+\frac{\omega_k+\omega_l}{2}} &
\frac{{\rm i}}{(\Omega^*_k-\Omega^*_l)+\frac{\omega_k+\omega_l}{2}}
 \end {array} \right),\\
&& G_{k,l}=\left( \begin {array}{cc}
\frac{{\rm i}}{(\Omega_k-\Omega_l)-\frac{\omega_k+\omega_l}{2}} \frac{\Omega_k-{\rm i}a - \frac{\omega_k}{2} }{\Omega_l-{\rm i}a + \frac{\omega_l}{2}} &
\frac{{\rm i}}{(\Omega_k+\Omega^*_l)-\frac{\omega_k+\omega_l}{2}} \frac{\Omega_k-{\rm i}a -\frac{\omega_k}{2} }{-\Omega^*_l-{\rm i}a +\frac{\omega_l}{2} }  \\
\frac{{\rm i}}{(\Omega^*_k+\Omega_l)+\frac{\omega_k+\omega_l}{2}} \frac{\Omega^*_k+{\rm i}a + \frac{\omega_k}{2} }{-\Omega_l+{\rm i}a - \frac{\omega_l}{2}  } &
\frac{{\rm i}}{(\Omega^*_k-\Omega^*_l)+\frac{\omega_k+\omega_l}{2}} \frac{\Omega^*_k+{\rm i}a + \frac{\omega_k}{2}}{\Omega^*_l+{\rm i}a -\frac{\omega_l}{2} }
 \end {array} \right),
\end{eqnarray*}
with
\begin{eqnarray*}
&& \zeta_{k} =\omega_k x
+ \left[ -\frac{4\sigma \rho^2\omega_k}{(2{\rm i}\Omega_k+2a)^2+\omega^2_k} +2{\rm i}\omega_k\Omega_k \right]y
+ [\frac{4\sigma \rho^2\omega_k}{(2{\rm i}\Omega_k+2a)^2+\omega^2_k}+2a\omega_k] t
+\xi_{k,0}+\eta_{k,0}.
\end{eqnarray*}
where $\Omega_k$ are complex parameters and $\omega_k,\xi_{k,0},\eta_{k,0}$ are real parameters for $k=1,2,\cdots,M$.
When these parameters satisfy the constraint conditions:
\begin{eqnarray}
\frac{2\sigma \rho^2}{(2{\rm i}\Omega_k+2a)^2+\omega^2_k} -{\rm i}\Omega_k =0,
\end{eqnarray}
the solutions (\ref{br2-01}) are breather solutions for the one-dimensional YO system.

For example, When $M=1$, the one-breather solutions for the two-dimensional YO system are written as
\begin{eqnarray}
&&\label{one-breather2-01} S=\rho{\rm e}^{-{\rm i}\alpha t}\frac{g}{f},\ \ L=\alpha+2\left( \ln f \right)_{xx},\\
&&\label{one-breather2-02} f=1+ {\rm e}^{\zeta_1} + {\rm e}^{\zeta^*_1} +A_{12}{\rm e}^{\zeta_1+\zeta^*_1},\\
&&\label{one-breather2-03} g=1+ K_1{\rm e}^{\zeta_1} + K_2{\rm e}^{\zeta^*_1} +A_{12}K_1K_2{\rm e}^{\zeta_1+\zeta^*_1},
\end{eqnarray}
with $ A_{12}=\frac{(\Omega_1+\Omega^*_1)^2}{[(\Omega_1+\Omega^*_1)^2-\omega^2_1]}$,
$K_1= \frac{2{\rm i}\Omega_1+2a -{\rm i}\omega_1}{2{\rm i}\Omega_1+2a +{\rm i}\omega_1 }$,
$K_2=\frac{2{\rm i}\Omega^*_1-2a +{\rm i}\omega_1}{2{\rm i}\Omega^*_1-2a -{\rm i}\omega_1 }$ and
$\zeta_{1} =\omega_1 x
+ \left[ -\frac{4\sigma \rho^2\omega_1}{(2{\rm i}\Omega_1+2a)^2+\omega^2_1} +2{\rm i}\omega_1\Omega_1 \right]y
+ [\frac{4\sigma \rho^2\omega_1}{(2{\rm i}\Omega_1+2a)^2+\omega^2_1}+2a\omega_1] t
+\zeta_{1,0}$.

Let $\Omega_1=\Omega_{1R}+{\rm i}\Omega_{1I}$, then $\zeta_1=\zeta_{1R}+{\rm i}\zeta_{1I}$,
\begin{eqnarray}
&&\zeta_{1R}= \omega_1 x
- 2\omega_1 \left[ \frac{2\sigma \rho^2\Theta_1 }{\Theta^2_1+64\Omega^2_{1R}(\Omega_{1I}-a)^2} +\Omega_{1I} \right]y
+ 2\omega_1[\frac{2\sigma \rho^2\Theta_1 }{\Theta^2_1+64\Omega^2_{1R}(\Omega_{1I}-a)^2}+a] t
+\zeta_{1,0},\\
&&\zeta_{1I}=
 \frac{32\sigma \rho^2\omega_1\Omega_{1R}(\Omega_{1I}-a)}{\Theta^2_1+64\Omega^2_{1R}(\Omega_{1I}-a)^2} (t-y) +2\omega_1\Omega_{1R}y,
\end{eqnarray}
with $\Theta_1=4(\Omega_{1I}-a)^2+\omega^2_1-4\Omega^2_{1R}$.

\begin{figure*}[!htbp]
\centering
{\includegraphics[height=1.2in,width=3in]{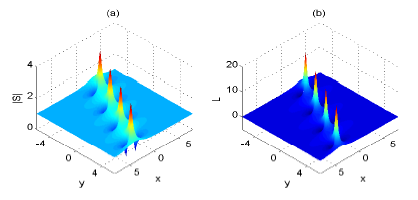}}
\caption{  The one-breather for two-dimensional YO system defined by Eqs. (\ref{one-breather2-01})--(\ref{one-breather2-03}) with the parameters $\sigma=\rho=a=\omega_1=1$, $\zeta_{2,0}=0$,  $\Omega_1=1+{\rm i}$ at $t=0$. \label{one-breather-2-2d}  }
\end{figure*}

It is straightforward to show that $S\rightarrow\rho{\rm e}^{-{\rm i}\alpha t}$, $L\rightarrow \alpha$ as $\zeta_{1R}\rightarrow -\infty$ and
$S\rightarrow\rho{\rm e}^{-{\rm i}\alpha t}\frac{2{\rm i}\Omega_1+2a -{\rm i}\omega_1}{2{\rm i}\Omega_1+2a +{\rm i}\omega_1 }\frac{2{\rm i}\Omega^*_1-2a +{\rm i}\omega_1}{2{\rm i}\Omega^*_1-2a -{\rm i}\omega_1 }$, $L\rightarrow \alpha$ as $\zeta_{1R}\rightarrow \infty$.
Thus this breather solution is localized along the direction of the line $\zeta_{1R}=0$ and periodic along the direction of the line $\zeta_{1I}=0$. This breather behavior is controlled by the complex parameter $\Omega_1$ and the real spatial frequency $\omega_1$.
Fig. \ref{one-breather-2-2d} illustrates the behavior of this breather solution only at the fixed time $t=0$.

\begin{figure*}[!htbp]
\centering
{\includegraphics[height=1.2in,width=3in]{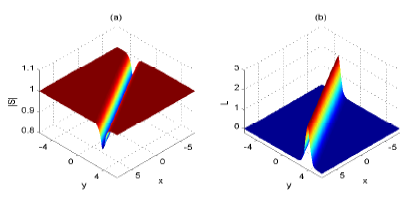}}
\caption{  The one-dark soliton for two-dimensional YO system defined by Eqs. (\ref{one-breather2-01})--(\ref{one-breather2-03}) and (\ref{one-breather2-dark-condition})  with the parameters $\sigma=\rho=a/3=\omega_1/2=1$, $\zeta_{2,0}=0$,  $\Omega_1={\rm i}$ at $t=0$.
\label{one-dark-2-2d} }
\end{figure*}

Note that if
\begin{eqnarray}\label{one-breather2-dark-condition}
\Omega_{1R}=0,
\end{eqnarray}
we find $\zeta_{1I}=0$ and then the breather solution reduces to dark soliton solution for the two-dimensional YO
system. One example of this solution, the dark soliton solution is shown in Fig.\ref{one-dark-2-2d}.

In general, letting $\Omega_k=\Omega_{kR}+{\rm i}\Omega_{kI}$ for $k=1,2,\cdots,M$, when the constraint conditions
\begin{eqnarray}\label{breather2-dark-condition}
\Omega_{kR}=0,
\end{eqnarray}
are satisfied, the multi-dark soliton solutions are constructed directly.

For the one-dimensional case, the constraint conditions
\begin{eqnarray}\label{breather2-1d-condition}
\Omega_{kR}=0,\ \ \Omega_{kI} + \frac{2\sigma\rho^2}{4(\Omega_{kI}-a)^2+\omega^2_k}=0
\end{eqnarray}
are satisfied, the multi-dark soliton solutions can be obtained.

\begin{figure*}[!htbp]
\centering
{\includegraphics[height=2.1in,width=4.3in]{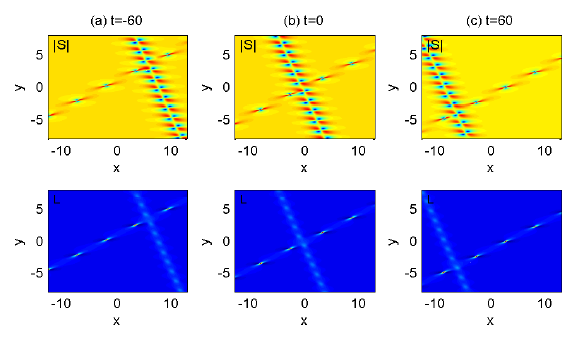}}
\caption{  The two-breather for two-dimensional YO system defined by Eqs.(\ref{breather-twosoliton-01})--(\ref{breather-twosoliton-03}) with the parameters $\sigma=\rho=\omega_1=1$, $\omega_2=\frac{6}{5}$, $a=\zeta_{2,0}=0$,  $\Omega_1=\frac{3}{2}+\frac{6}{5}{\rm i}$ and $\Omega_2=\frac{2}{3}-\frac{1}{3}{\rm i}$. }
\end{figure*}

\begin{figure*}[!htbp]
\centering
{\includegraphics[height=2.1in,width=4.3in]{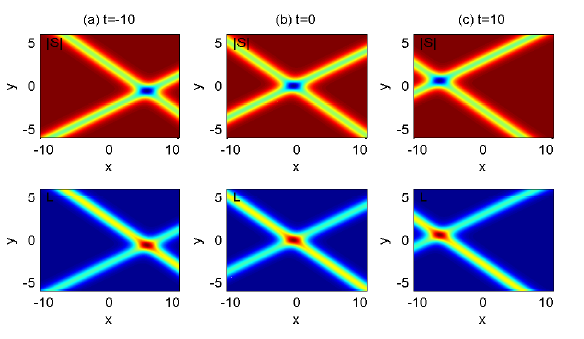}}
\caption{  The two-dark soliton for two-dimensional YO system defined by Eqs.(\ref{breather-twosoliton-01})--(\ref{breather-twosoliton-03}) and (\ref{breather2-dark-condition}) ($k=1,2$) with the parameters $\sigma=\rho=\omega_1=1$, $\omega_2=\frac{6}{5}$, $a=\zeta_{2,0}=0$,  $\Omega_1={\rm i}$ and $\Omega_2=-\frac{5}{4}{\rm i}$. }
\end{figure*}

\begin{figure*}[!htbp]
\centering
{\includegraphics[height=2.1in,width=4.3in]{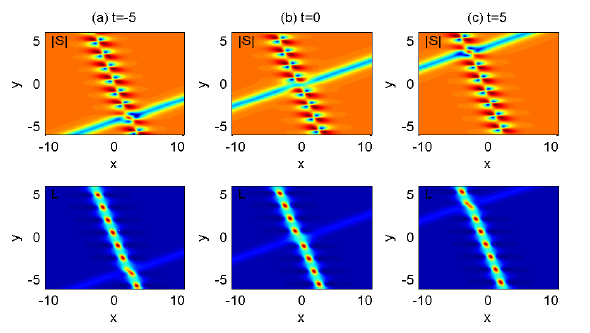}}
\caption{The mixed solution consisting of one breather and one dark soliton for two-dimensional YO system defined
by Eqs. (\ref{breather-twosoliton-01})--(\ref{breather-twosoliton-03}) and (\ref{breather2-dark-condition}) ($k=1$)  with the parameters $\sigma=\rho=\omega_1=1$, $\omega_2=\frac{6}{5}$, $a=\zeta_{2,0}=0$,  $\Omega_1=\frac{1}{5}{\rm i}$ and $\Omega_2=\frac{2}{3}-\frac{1}{3}{\rm i}$. }
\end{figure*}

When $M=2$, the two-breather solutions for the two-dimensional YO system are written as
\begin{eqnarray}
&&\label{breather-twosoliton-01} S=\rho{\rm e}^{-{\rm i}\alpha t}\frac{G}{F},\ \ L=\alpha+2\left( \ln \Delta_0 F \right)_{xx},\\
&&\label{breather-twosoliton-02} F=\left| \begin {array}{cccc}
-\frac{{\rm i}}{\omega_1 {\rm e}^{\zeta_1}} - \frac{{\rm i}}{\omega_1} &
\frac{{\rm i} }{\Omega_1+\Omega^*_1-\omega_1} &
\frac{{\rm i}}{(\Omega_1-\Omega_2)-\frac{\omega_1+\omega_2}{2}} &
\frac{{\rm i}}{(\Omega_1+\Omega^*_2)-\frac{\omega_1+\omega_2}{2}}\\
 \frac{ {\rm i} }{\Omega_1+\Omega^*_1+\omega_1}  &
 \frac{{\rm i}}{\omega_1 {\rm e}^{\zeta^*_1}} + \frac{{\rm i}}{\omega_1}  &
 \frac{{\rm i}}{(\Omega^*_1+\Omega_2)+\frac{\omega_1+\omega_2}{2}} &
\frac{{\rm i}}{(\Omega^*_1-\Omega^*_2)+\frac{\omega_1+\omega_2}{2}}\\
\frac{{\rm i}}{(\Omega_2-\Omega_1)-\frac{\omega_2+\omega_1}{2}} &
\frac{{\rm i}}{(\Omega_2+\Omega^*_1)-\frac{\omega_2+\omega_1}{2}} &
-\frac{{\rm i}}{\omega_2 {\rm e}^{\zeta_2}} - \frac{{\rm i}}{\omega_2} &
\frac{{\rm i} }{\Omega_2+\Omega^*_2-\omega_2} \\
\frac{{\rm i}}{(\Omega^*_2+\Omega_1)+\frac{\omega_2+\omega_1}{2}} &
\frac{{\rm i}}{(\Omega^*_2-\Omega^*_1)+\frac{\omega_2+\omega_1}{2}} &
\frac{ {\rm i} }{\Omega_2+\Omega^*_2+\omega_2}  &
 \frac{{\rm i}}{\omega_2 {\rm e}^{\zeta^*_2}} + \frac{{\rm i}}{\omega_2}
 \end {array} \right|,\\
&&\label{breather-twosoliton-03}
 \hspace{-1cm} G=\left| \begin {array}{cccc}
 -\frac{{\rm i}}{\omega_1 {\rm e}^{\zeta_1}} - \frac{{\rm i}}{\omega_1}\frac{\Omega_1-{\rm i}a -\frac{\omega_1}{2}}{\Omega_1-{\rm i}a +\frac{\omega_1}{2}} &
\frac{ {\rm i} }{\Omega_1+\Omega^*_1-\omega_1} \frac{\Omega_1-{\rm i}a -\frac{\omega_1}{2} }{ -\Omega^*_1-{\rm i}a +\frac{\omega_1}{2} } &
 \frac{{\rm i}}{(\Omega_1-\Omega_2)-\frac{\omega_1+\omega_2}{2}} \frac{\Omega_1-{\rm i}a - \frac{\omega_1}{2} }{\Omega_2-{\rm i}a + \frac{\omega_2}{2}} &
\frac{{\rm i}}{(\Omega_1+\Omega^*_2)-\frac{\omega_1+\omega_2}{2}} \frac{\Omega_1-{\rm i}a -\frac{\omega_1}{2} }{-\Omega^*_2-{\rm i}a +\frac{\omega_2}{2} } \\
 \frac{{\rm i} }{\Omega_1+\Omega^*_1+\omega_1} \frac{\Omega^*_1+{\rm i}a + \frac{\omega_1}{2}}{-\Omega_1+{\rm i}a - \frac{\omega_1}{2}} &
 \frac{{\rm i}}{\omega_1 {\rm e}^{\zeta^*_1}} + \frac{{\rm i}}{\omega_1} \frac{\Omega^*_1+{\rm i}a +\frac{\omega_1}{2}}{\Omega^*_1+{\rm i}a -\frac{\omega_1}{2} } &
 \frac{{\rm i}}{(\Omega^*_1+\Omega_2)+\frac{\omega_1+\omega_2}{2}} \frac{\Omega^*_1+{\rm i}a + \frac{\omega_1}{2} }{-\Omega_2+{\rm i}a - \frac{\omega_2}{2}  } &
\frac{{\rm i}}{(\Omega^*_1-\Omega^*_2)+\frac{\omega_1+\omega_2}{2}} \frac{\Omega^*_1+{\rm i}a + \frac{\omega_1}{2}}{\Omega^*_2+{\rm i}a -\frac{\omega_2}{2} }\\
\frac{{\rm i}}{(\Omega_2-\Omega_1)-\frac{\omega_2+\omega_1}{2}} \frac{\Omega_2-{\rm i}a - \frac{\omega_2}{2} }{\Omega_1-{\rm i}a + \frac{\omega_1}{2}} &
\frac{{\rm i}}{(\Omega_2+\Omega^*_1)-\frac{\omega_2+\omega_1}{2}} \frac{\Omega_2-{\rm i}a -\frac{\omega_2}{2} }{-\Omega^*_1-{\rm i}a +\frac{\omega_1}{2} } &
-\frac{{\rm i}}{\omega_2 {\rm e}^{\zeta_2}} - \frac{{\rm i}}{\omega_2}\frac{\Omega_2-{\rm i}a -\frac{\omega_2}{2}}{\Omega_2-{\rm i}a +\frac{\omega_2}{2}} &
\frac{ {\rm i} }{\Omega_2+\Omega^*_2-\omega_2} \frac{\Omega_2-{\rm i}a -\frac{\omega_2}{2} }{ -\Omega^*_2-{\rm i}a +\frac{\omega_2}{2} }  \\
\frac{{\rm i}}{(\Omega^*_2+\Omega_1)+\frac{\omega_2+\omega_1}{2}} \frac{\Omega^*_2+{\rm i}a + \frac{\omega_2}{2} }{-\Omega_1+{\rm i}a - \frac{\omega_1}{2}  } &
\frac{{\rm i}}{(\Omega^*_2-\Omega^*_1)+\frac{\omega_2+\omega_1}{2}} \frac{\Omega^*_2+{\rm i}a + \frac{\omega_2}{2}}{\Omega^*_1+{\rm i}a -\frac{\omega_1}{2} } &
\frac{{\rm i} }{\Omega_2+\Omega^*_2+\omega_2} \frac{\Omega^*_2+{\rm i}a + \frac{\omega_2}{2}}{-\Omega_2+{\rm i}a - \frac{\omega_2}{2}} &
 \frac{{\rm i}}{\omega_2 {\rm e}^{\zeta^*_2}} + \frac{{\rm i}}{\omega_2} \frac{\Omega^*_2+{\rm i}a +\frac{\omega_2}{2}}{\Omega^*_2+{\rm i}a -\frac{\omega_2}{2} }
 \end {array} \right|,\ \ \ \ \ \ \ \ \
\end{eqnarray}
where $\Delta_0=\omega^2_1\omega^2_2\textmd{e}^{\zeta_1+\zeta^*_1+\zeta_2+\zeta^*_2} $ and
\begin{eqnarray*}
&& \zeta_{k} =\omega_k x
+ \left[ -\frac{4\sigma \rho^2\omega_k}{(2{\rm i}\Omega_k+2a)^2+\omega^2_k} +2{\rm i}\omega_k\Omega_k \right]y
+ [\frac{4\sigma \rho^2\omega_k}{(2{\rm i}\Omega_k+2a)^2+\omega^2_k}+2a\omega_k] t
+\xi_{k,0}+\eta_{k,0}.
\end{eqnarray*}
for $k=1,2$.

This solution contains the two breather solution, two-dark soliton solution and the mixed
solution consisting of one breather and one dark soliton  for the two-dimensional YO system, which are shown in Fig.18-20.

\subsection{Rational solution}
 Because this result is equivalent to the result in Section 3.2, we don't list it here again.

\subsection{Exp-rational breather }


By using the same procedure as in Section 3.3, we have the following exp-rational solutions.

\textbf{Theorem 4.2} The $\tilde{N}$-rational-$\tilde{N}'$-exp solutions for two-dimensional YO system are
\begin{eqnarray}
 S=\rho{\rm e}^{-{\rm i}\alpha t}\frac{g}{f},\ \ L=\alpha+2\left( \ln f \right)_{xx},
\end{eqnarray}
where $f= \Delta_0|F_{k,l}|$, $g= \Delta_0|G_{k,l}|$ and $\Delta_0= \textmd{e}^{\sum^{s'_{\tilde{N}'}}_{k=s'_1}\zeta_k+\zeta^*_k} \prod^{s'_{\tilde{N}'}}_{k=s'_1}\omega^2_k$, and the matrix elements are defined as follows:

\begin{eqnarray}
 |F_{k,l}|=\left|  \begin {array}{cc} A & B\\ C & D \end {array} \right|,\ \
|G_{k,l}|=\left|  \begin {array}{cc} \mathcal{A} & \mathcal{B}\\ \mathcal{C} & \mathcal{D} \end {array} \right|,
\end{eqnarray}
where $A$ and $\mathcal{A}$ are $2\tilde{N}\times 2\tilde{N}$ matrices defined by
\begin{eqnarray*}
&& A_{k,k}= \left( \begin {array}{cc}
\theta_k &  \frac{{\rm i}}{\Omega_k+\Omega^*_k} \\
 \frac{{\rm i}}{\Omega_k+\Omega^*_k}  &  \theta^*_k
 \end {array} \right),\ \
 \mathcal{A}_{k,k}= \left( \begin {array}{cc}
\theta_k-\frac{1}{{\rm i}\Omega_k+a} &
   \frac{{\rm i}}{\Omega_k+\Omega^*_k}\frac{-\Omega_k+{\rm i}a}{\Omega^*_k+{\rm i}a} \\
   \frac{{\rm i}}{\Omega_k+\Omega^*_k} \frac{-\Omega^*_k-{\rm i}a}{\Omega_k-{\rm i}a}  &
 \theta^*_k-\frac{1}{{\rm i}\Omega^*_k-a}
 \end {array} \right),
\end{eqnarray*}
\begin{eqnarray*}
&& A_{k,l}=\left( \begin {array}{cc}
\frac{{\rm i}}{\Omega_k-\Omega_l} &
\frac{{\rm i}}{\Omega_k+\Omega^*_l}\\
\frac{{\rm i}}{\Omega^*_k-\Omega_l} &
\frac{{\rm i}}{\Omega^*_k-\Omega^*_l}
 \end {array} \right),\ \
 \mathcal{A}_{k,l}=\left( \begin {array}{cc}
\frac{{\rm i}}{\Omega_k-\Omega_l}
\frac{ {\rm i}\Omega_k +a   }{ {\rm i}\Omega_l +a  } &
\frac{{\rm i}}{\Omega_k+\Omega^*_l}
\frac{ {\rm i}\Omega_k +a   }{ -{\rm i}\Omega^*_l +a } \\
\frac{{\rm i}}{\Omega^*_k-\Omega_l}
\frac{ -{\rm i}\Omega^*_k +a  }{ {\rm i}\Omega_l +a   } &
\frac{{\rm i}}{\Omega^*_k-\Omega^*_l}
\frac{ -{\rm i}\Omega^*_k +a }{ -{\rm i}\Omega^*_l +a }
 \end {array} \right),\\
\end{eqnarray*}
$D$ and $\mathcal{D}$ are $2\tilde{N}'\times 2\tilde{N}'$ matrices defined by
\begin{eqnarray*}
&& D_{k,k}= \left( \begin {array}{cc}
-\frac{{\rm i}}{\omega_k {\rm e}^{\zeta_k}} - \frac{{\rm i}}{\omega_k} &
\frac{{\rm i} }{\Omega_k+\Omega^*_k-\omega_k} \\
 \frac{ {\rm i} }{\Omega_k+\Omega^*_k+\omega_k}  &
 \frac{{\rm i}}{\omega_k {\rm e}^{\zeta^*_k}} + \frac{{\rm i}}{\omega_k}
 \end {array} \right),\\
&& \mathcal{D}_{k,k}= \left( \begin {array}{cc}
-\frac{{\rm i}}{\omega_k {\rm e}^{\zeta_k}} - \frac{{\rm i}}{\omega_k}\frac{\Omega_k-{\rm i}a -\frac{\omega_k}{2}}{\Omega_k-{\rm i}a +\frac{\omega_k}{2}} &
\frac{ {\rm i} }{\Omega_k+\Omega^*_k-\omega_k} \frac{\Omega_k-{\rm i}a -\frac{\omega_k}{2} }{ -\Omega^*_k-{\rm i}a +\frac{\omega_k}{2} }\\
 \frac{{\rm i} }{\Omega_k+\Omega^*_k+\omega_k} \frac{\Omega^*_k+{\rm i}a + \frac{\omega_k}{2}}{-\Omega_k+{\rm i}a - \frac{\omega_k}{2}} &
 \frac{{\rm i}}{\omega_k {\rm e}^{\zeta^*_k}} + \frac{{\rm i}}{\omega_k} \frac{\Omega^*_k+{\rm i}a +\frac{\omega_k}{2}}{\Omega^*_k+{\rm i}a -\frac{\omega_k}{2} }
 \end {array} \right),\\
&& D_{k,l}=\left( \begin {array}{cc}
\frac{{\rm i}}{(\Omega_k-\Omega_l)-\frac{ \omega_k+  \omega_l}{2}} &
\frac{{\rm i}}{(\Omega_k+\Omega^*_l)-\frac{\omega_k+ \omega_l}{2}}\\
\frac{{\rm i}}{(\Omega^*_k-\Omega_l)+\frac{ \omega_k+ \omega_l}{2}} &
\frac{{\rm i}}{(\Omega^*_k-\Omega^*_l)+\frac{\omega_k+ \omega_l}{2}}
 \end {array} \right),\\
&& \mathcal{D}_{k,l}=\left( \begin {array}{cc}
\frac{{\rm i}}{(\Omega_k-\Omega_l)-\frac{\omega_k+  \omega_l}{2}}
\frac{ {\rm i}\Omega_k +a -{\rm i} \frac{\omega_k}{2}  }{ {\rm i}\Omega_l +a + {\rm i} \frac{\omega_l}{2}  } &
\frac{{\rm i}}{(\Omega_k+\Omega^*_l)-\frac{\omega_k+ \omega_l}{2}}
\frac{ {\rm i}\Omega_k +a - {\rm i} \frac{\omega_k}{2}  }{ -{\rm i}\Omega^*_l +a + {\rm i} \frac{\omega_l}{2}  } \\
\frac{{\rm i}}{(\Omega^*_k-\Omega_l)+\frac{ \omega_k+ \omega_l}{2}}
\frac{ -{\rm i}\Omega^*_k +a - {\rm i} \frac{\omega_k}{2}  }{ {\rm i}\Omega_l +a +{\rm i} \frac{\omega_l}{2}  } &
\frac{{\rm i}}{(\Omega^*_k-\Omega^*_l)+\frac{ \omega_k+ \omega_l}{2}}
\frac{ -{\rm i}\Omega^*_k +a - {\rm i} \frac{\omega_k}{2}  }{ -{\rm i}\Omega^*_l +a + {\rm i} \frac{\omega_l}{2}  }
 \end {array} \right),\\
\end{eqnarray*}
$B$ and $\mathcal{B}$ are $2\tilde{N}\times 2\tilde{N}'$ matrices defined by
\begin{eqnarray*}
&& B_{k,l}=\left( \begin {array}{cc}
\frac{{\rm i}}{(\Omega_k-\Omega_l)-\frac{ \omega_l}{2}} &
\frac{{\rm i}}{(\Omega_k+\Omega^*_l)-\frac{ \omega_l}{2}}\\
\frac{{\rm i}}{(\Omega^*_k-\Omega_l)+\frac{\omega_l}{2}} &
\frac{{\rm i}}{(\Omega^*_k-\Omega^*_l)+\frac{\omega_l}{2}}
 \end {array} \right),\\
&& \mathcal{B}_{k,l}=\left( \begin {array}{cc}
\frac{{\rm i}}{(\Omega_k-\Omega_l)-\frac{ \omega_l}{2}}
\frac{ {\rm i}\Omega_k +a  }{ {\rm i}\Omega_l +a + {\rm i} \frac{\omega_l}{2}  } &
\frac{{\rm i}}{(\Omega_k+\Omega^*_l)-\frac{ \omega_l}{2}}
\frac{ {\rm i}\Omega_k +a   }{ -{\rm i}\Omega^*_l +a + {\rm i} \frac{\omega_l}{2}  } \\
\frac{{\rm i}}{(\Omega^*_k-\Omega_l)+\frac{ \omega_l}{2}}
\frac{ -{\rm i}\Omega^*_k +a }{ {\rm i}\Omega_l +a + {\rm i} \frac{\omega_l}{2}  } &
\frac{{\rm i}}{(\Omega^*_k-\Omega^*_l)+\frac{ \omega_l}{2}}
\frac{ -{\rm i}\Omega^*_k +a }{ -{\rm i}\Omega^*_l +a + {\rm i} \frac{\omega_l}{2}  }
 \end {array} \right),
\end{eqnarray*}
$C$ and $\mathcal{C}$ are $2\tilde{N}'\times 2\tilde{N}$ matrices defined by
\begin{eqnarray*}
&& C_{k,l}=\left( \begin {array}{cc}
\frac{{\rm i}}{(\Omega_k-\Omega_l)-\frac{\omega_k}{2}} &
\frac{{\rm i}}{(\Omega_k+\Omega^*_l)-\frac{\omega_k}{2}}\\
\frac{{\rm i}}{(\Omega^*_k-\Omega_l)+\frac{\omega_k}{2}} &
\frac{{\rm i}}{(\Omega^*_k-\Omega^*_l)+\frac{\omega_k}{2}}
 \end {array} \right),\\
&& C_{k,l}=\left( \begin {array}{cc}
\frac{{\rm i}}{(\Omega_k-\Omega_l)-\frac{ \omega_k}{2}}
\frac{ {\rm i}\Omega_k +a - {\rm i} \frac{\omega_k}{2}  }{ {\rm i}\Omega_l +a   } &
\frac{{\rm i}}{(\Omega_k+\Omega^*_l)-\frac{\omega_k}{2}}
\frac{ {\rm i}\Omega_k +a - {\rm i} \frac{\omega_k}{2}  }{ -{\rm i}\Omega^*_l +a  } \\
\frac{{\rm i}}{(\Omega^*_k-\Omega_l)+\frac{ \omega_k}{2}}
\frac{ -{\rm i}\Omega^*_k +a - {\rm i} \frac{\omega_k}{2}  }{ {\rm i}\Omega_l +a  } &
\frac{{\rm i}}{(\Omega^*_k-\Omega^*_l)+\frac{\omega_k}{2}}
\frac{ -{\rm i}\Omega^*_k +a - {\rm i} \frac{\omega_k}{2}  }{ -{\rm i}\Omega^*_l +a }
 \end {array} \right),\\
\end{eqnarray*}
and
\begin{eqnarray*}
&& \theta_k=-{\rm i}x + \left[\frac{{\rm i}\sigma\rho^2}{({\rm i}\Omega_k+a)^2} +2\Omega_k \right]y - \left[\frac{{\rm i}\sigma\rho^2}{({\rm i}\Omega_k+a)^2} +2{\rm i}a \right]t,\\
&&  \zeta_{k} =\omega_k x
+ \left[ -\frac{4\sigma \rho^2\omega_k}{(2{\rm i}\Omega_k+2a)^2+\omega^2_k} +2{\rm i}\omega_k\Omega_k \right]y
+ [\frac{4\sigma \rho^2\omega_k}{(2{\rm i}\Omega_k+2a)^2+\omega^2_k}+2a\omega_k] t
+\xi_{k,0}+\eta_{k,0}.
\end{eqnarray*}

When two-dimensional YO system reduces to one-dimensional case, one only obtain one-rational-$\tilde{N}'$-exp solution (one-rogue-wave-$\tilde{N}'$-breather solution) with the following constraints conditions:
\begin{eqnarray}
&& \frac{{\rm i}\sigma\rho^2}{({\rm i}\Omega_1+a)^2} +2\Omega_1=0,\\
&& -\frac{4\sigma \rho^2\omega_k}{(2{\rm i}\Omega_k+2a)^2+\omega^2_k} +2{\rm i}\omega_k\Omega_k=0.
\end{eqnarray}

\begin{figure*}[!htbp]
\centering
{\includegraphics[height=2.1in,width=6in]{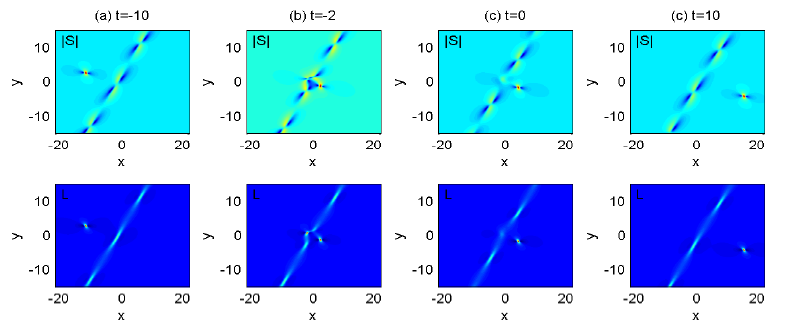}}
\caption{The mixed solution consisting of the breather and one lump for two-dimensional YO system defined
by Eqs. (\ref{one-rational-one-exp2-01})--(\ref{one-rational-one-exp2-03}) with the parameters $\sigma=\rho=\omega_2=1$, $a=\zeta_{2,0}=0$, $\Omega_1=\frac{\sqrt{7}}{4}+\frac{1}{4}{\rm i}$ and $\Omega_2=\frac{2}{3}+\frac{2}{5}{\rm i}$. }
\end{figure*}

\begin{figure*}[!htbp]
\centering
{\includegraphics[height=2.1in,width=6in]{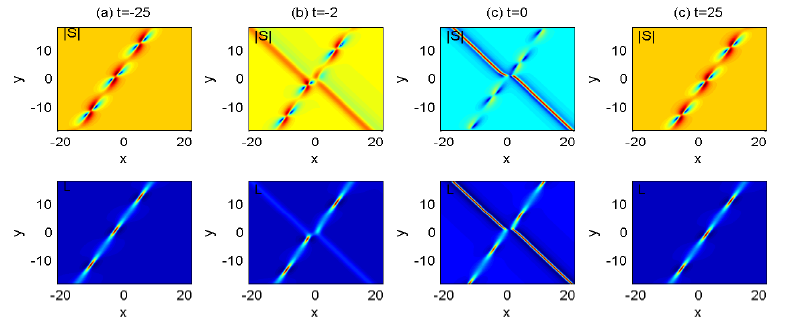}}
\caption{ The mixed solution consisting of the breather and rogue wave for two-dimensional YO system defined
by Eqs. (\ref{one-rational-one-exp2-01})--(\ref{one-rational-one-exp2-03}) and (\ref{byo-61}) with the parameters $\sigma=\rho=\omega_2=1$, $a=\zeta_{2,0}=0$,  $\Omega_1=1+\frac{1}{5}{\rm i}$ and $\Omega_2=\frac{2}{3}+\frac{2}{5}{\rm i}$. }
\end{figure*}

\begin{figure*}[!htbp]
\centering
{\includegraphics[height=2.1in,width=6in]{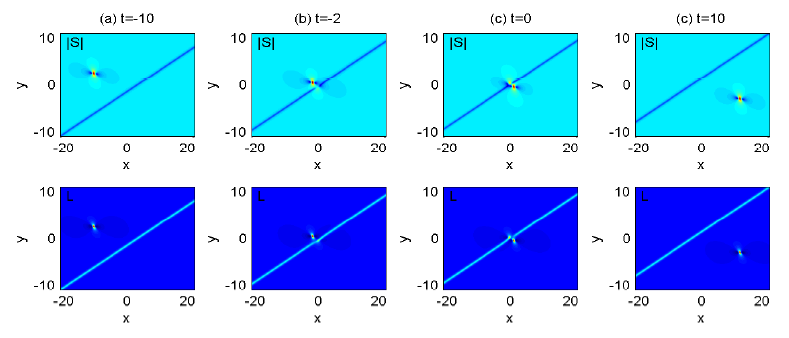}}
\caption{The mixed solution consisting of one dark soliton and one lump orbit for two-dimensional YO system defined
by Eqs. (\ref{one-rational-one-exp2-01})--(\ref{one-rational-one-exp2-03}) and (\ref{byo-61}), (\ref{breather2-dark-condition})($k=2$) with the parameters $\sigma=\rho=\omega_2/3=1$, $a=\zeta_{2,0}=0$,  $\Omega_1=1+\frac{1}{5}{\rm i}$ and $\Omega_2={\rm i}$. }
\end{figure*}

For example, when $\tilde{N}=\tilde{N}'=1$, one-rational-one-exp solution is given by
\begin{eqnarray}
&&\label{one-rational-one-exp2-01} S=\rho{\rm e}^{-{\rm i}\alpha t}\frac{G}{F},\ \ L=\alpha+2\left( \ln \Delta_0 F \right)_{xx},\\
&&\label{one-rational-one-exp2-02}
 F=\left| \begin {array}{cccc} \theta_1
&\frac{{\rm i}}{\Omega^*_1+\Omega_1}
&\frac{{\rm i}}{\Omega_1-\Omega_2 -\frac{\omega_2}{2}}
&\frac{{\rm i}}{\Omega^*_2+\Omega_1 -\frac{\omega_2}{2}} \\
\frac{{\rm i}}{\Omega^*_1+\Omega_1}
&\theta^*_1
&\frac{{\rm i}}{\Omega^*_1+\Omega_2 +\frac{\omega_2}{2}}
&\frac{{\rm i}}{\Omega^*_1-\Omega^*_2 +\frac{\omega_2}{2}} \\
\frac{{\rm i}}{\Omega_2-\Omega_1 -\frac{\omega_2}{2}}
& \frac{{\rm i}}{\Omega^*_1+\Omega_2 -\frac{\omega_2}{2}}
& -\frac{{\rm i}}{ \omega_2 {\rm e}^{\zeta_2} } -\frac{{\rm i}}{\omega_2}
&  \frac{{\rm i}}{\Omega^*_2+\Omega_2-\omega_2} \\
\frac{{\rm i}}{\Omega^*_2+\Omega_1 +\frac{\omega_2}{2}}
&\frac{{\rm i}}{\Omega^*_2-\Omega^*_1 +\frac{\omega_2}{2}}
& \frac{{\rm i}}{\Omega^*_2+\Omega_2+\omega_2}
& \frac{{\rm i}}{ \omega_2 {\rm e}^{\zeta^*_2} } +\frac{{\rm i}}{\omega_2}  \end {array} \right|,\\
 &&\label{one-rational-one-exp2-03}
G=\left| \begin {array}{cccc} \theta_1-\frac{1}{{\rm i}\Omega_1+a} &
\frac{-{\rm i}}{\Omega^*_1+\Omega_1} \frac{\Omega_1-{\rm i}a}{\Omega^*_1+{\rm i}a}
&\frac{{\rm i}}{\Omega_2-\Omega_1 -\frac{\omega_2}{2}}
\frac{{\rm i}\Omega_1+a}{{\rm i}\Omega_2+a+{\rm i}\frac{\omega_2}{2}}
&\frac{{\rm i}}{\Omega^*_2+\Omega_1 -\frac{\omega_2}{2}}\frac{{\rm i}\Omega_1+a}{-{\rm i}\Omega^*_2+a+{\rm i}\frac{\omega_2}{2}} \\
\frac{-{\rm i}}{\Omega^*_1+\Omega_1} \frac{\Omega^*_1+{\rm i}a}{\Omega_1-{\rm i}a} &
\theta^*_1 -\frac{1}{{\rm i}\Omega^*_1-a} &
\frac{{\rm i}}{\Omega^*_1-\Omega_2 +\frac{\omega_2}{2}} \frac{-{\rm i}\Omega^*_1+a}{{\rm i}\Omega_2+a+{\rm i}\frac{\omega_2}{2}} &
\frac{{\rm i}}{\Omega^*_1-\Omega^*_2 +\frac{\omega_2}{2}} \frac{-{\rm i}\Omega^*_1+a}{-{\rm i}\Omega^*_2+a+{\rm i}\frac{\omega_2}{2}} \\
\frac{{\rm i}}{\Omega_1-\Omega_2 +\frac{\omega_2}{2}} \frac{{\rm i}\Omega_2+a-{\rm i}\frac{\omega_2}{2}}{{\rm i}\Omega_1+a} &
\frac{{\rm i}}{\Omega^*_1-\Omega_2 +\frac{\omega_2}{2}} \frac{{\rm i}\Omega_2+a-{\rm i}\frac{\omega_2}{2}}{-{\rm i}\Omega^*_1+a} &
-\frac{{\rm i}}{ \omega_2 {\rm e}^{\zeta_2} } -\frac{{\rm i}}{\omega_2}
\frac{\Omega_2-{\rm i}a-\frac{\omega_2}{2}}{\Omega_2-{\rm i}a+\frac{\omega_2}{2}}
& \frac{-{\rm i}}{\Omega^*_2+\Omega_2-\omega_2}\frac{\Omega_2-{\rm i}a-\frac{\omega_2}{2}}{\Omega^*_2+{\rm i}a-\frac{\omega_2}{2}} \\
\frac{{\rm i}}{\Omega^*_2-\Omega_1 +\frac{\omega_2}{2}} \frac{-{\rm i}\Omega^*_2+a-{\rm i}\frac{\omega_2}{2}}{{\rm i}\Omega_1+a} &
\frac{{\rm i}}{\Omega^*_2-\Omega^*_1 +\frac{\omega_2}{2}} \frac{-{\rm i}\Omega^*_2+a-{\rm i}\frac{\omega_2}{2}}{-{\rm i}\Omega^*_1+a} &
\frac{-{\rm i}}{\Omega^*_2+\Omega_2+\omega_2}\frac{\Omega^*_2+{\rm i}a+\frac{\omega_2}{2}}{\Omega_2-{\rm i}a+\frac{\omega_2}{2}} &
\frac{{\rm i}}{ \omega_2 {\rm e}^{\zeta^*_2} } +\frac{{\rm i}}{\omega_2}\frac{\Omega^*_2+{\rm i}a+\frac{\omega_2}{2}}{\Omega^*_2+{\rm i}a-\frac{\omega_2}{2}}  \end {array} \right|,
\end{eqnarray}
where $\Delta_0=\omega^2_2  {\rm e}^{\zeta_2+\zeta^*_2}$ and
\begin{eqnarray*}
&& \theta_1=-{\rm i}x + \left[\frac{{\rm i}\sigma\rho^2}{({\rm i}\Omega_1+a)^2} +2\Omega_1 \right]y - \left[\frac{{\rm i}\sigma\rho^2}{({\rm i}\Omega_1+a)^2} +2{\rm i}a \right]t,\\
&&  \zeta_{2} =\omega_2 x
+ \left[ -\frac{4\sigma \rho^2\omega_2}{(2{\rm i}\Omega_2+2a)^2+\omega^2_2} +2{\rm i}\omega_2\Omega_2 \right]y
+ [\frac{4\sigma \rho^2\omega_2}{(2{\rm i}\Omega_2+2a)^2+\omega^2_2}+2a\omega_2] t
+\zeta_{2,0}.
\end{eqnarray*}

This solution represents the mixed solution consisting of one lump (and one rogue wave) and one general breather
(and one soliton). Examples of these mixed solutions are shown in Fig.21-24.
Here, we also present one example of the mixed solution consisting of one rogue wave and one soliton
for one-dimensional YO system as in Fig.25.

\begin{figure*}[!htbp]
\centering
{\includegraphics[height=2.1in,width=6in]{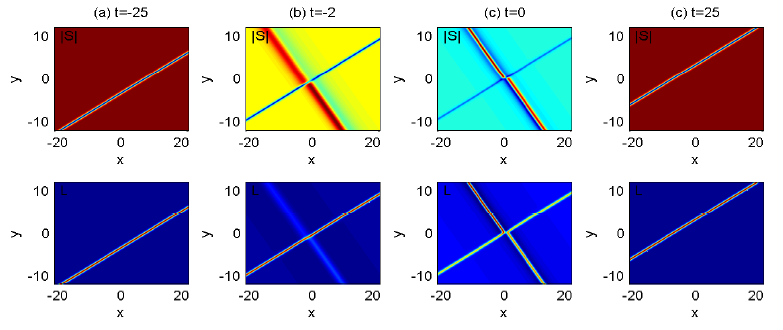}}
\caption{ The mixed solution consisting of one dark soliton and rogue wave for two-dimensional YO system defined
by Eqs. (\ref{one-rational-one-exp2-01})--(\ref{one-rational-one-exp2-03}) and (\ref{breather2-dark-condition})($k=2$) with the parameters $\sigma=\rho=\omega_2/3=1$, $a=\zeta_{2,0}=0$,  $\Omega_1=\frac{\sqrt{7}}{4}+\frac{1}{4}{\rm i}$ and $\Omega_2={\rm i}$. }
\end{figure*}

\begin{figure*}[!htbp]
\centering
{\includegraphics[height=1.2in,width=3in]{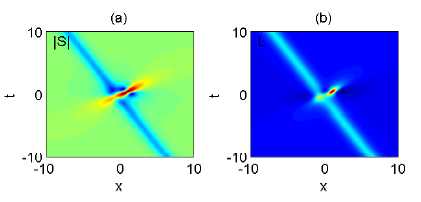}}
\caption{  The mixed solution consisting of one rogue wave and one soliton for one-dimensional YO system
defined by Eqs. (\ref{one-rational-one-exp2-01})--(\ref{one-rational-one-exp2-03}) and (\ref{breather2-1d-condition})($k=1$)  with the parameters $\sigma=\rho=\omega_2/3=1$, $a=\zeta_{2,0}=0$,  $\Omega_1=\frac{\sqrt[3]{4}}{4}(\sqrt{3}+{\rm i})$ and $\Omega_2=(-\frac{\sqrt[3]{54+3\sqrt{327}}}{6}+\frac{1}{2\sqrt[3]{54+3\sqrt{327}}}){\rm i}$. }
\end{figure*}

\section{Summary and discussions}

In this paper, by using bilinear method and the KP hierarchy reduction method, we construct the breather solutions for the YO system in one- and two-dimensional cases.
Similar to Akhmediev and Kuznetsov-Ma breather solutions (the wavenumber $k_i\rightarrow {\rm i}k_i$ ) for the nonlinear Schr\"{o}dinger equation,
is shown that the YO system have two kinds of breather solutions with the relations
$p_{2k-1}\rightarrow{\rm i}p_{2k-1}$, $p_{2k}\rightarrow-{\rm i}p_{2k}$, $q_{2k-1}\rightarrow{\rm i}q_{2k-1}$ and $q_{2k}\rightarrow-{\rm i}q_{2k}$, in which the homoclinic orbit and dark soliton solutions are two special cases respectively.
Furthermore, taking the long wave limit, we derive the rational and rational and rational-exp solutions
which contain lump, line rogue wave, soliton and their mixed cases.
By considering the further reduction, such solutions can be reduced to one-dimensional YO system.

\section{Acknowledgment}

JC and YC's work was supported from National
Natural Science Foundation of China (Nos. 11675054, 11705077 and 11775104), and Shanghai Collaborative
Innovation Center of Trustworthy Software for Internet of Things (No. ZF1213). BF's work was partially supported
by National Natural Science Foundation (DMS-1715991) and by the COS Research Enhancement Seed Grants
Program at UTRGV. KM's work was supported by JSPS Grant-in-Aid for Scientific Research (C-15K04909) and
JST CREST.








\end{document}